\documentclass[12pt]{JHEP3}
\usepackage{epsfig}
\usepackage{amsmath}
\newcommand{\p}{\partial}
\title{
Imaginary Sources: Completeness Conjecture and Charges}
\author{
Koji Hashimoto$^{\dagger}$ and John E. Wang$^{*}$\\
${}^\dagger$ {\it Institute of Physics, University of Tokyo, Komaba,
Tokyo, Japan}\\
\hspace{2mm} {\it Kavli Institute for Theoretical Physics,
University of California, Santa Barbara}\\
\hspace{2mm} {\it DAMTP, CMS, Cambridge University,
Wilberforce Road, Cambridge, UK}\\
\hspace{2mm} E-mail: \email{koji@hep1.c.u-tokyo.ac.jp}\\
${}^*$ {\it Physics Division, National Center for Theoretical
Sciences, Hsinchu, Taiwan}\\
\hspace{2mm} {\it Department of Physics, National Taiwan University,
Taipei, Taiwan}\\
\hspace{2mm} E-mail: \email{jwang@phys.cts.nthu.edu.tw, hllywd2@phys.ntu.edu.tw}\\
}

\abstract{While in string theory the subject of sources in
imaginary time has received some attention, we demonstrate the
power of imaginary sources by proving that they constitute in
several field theories a complete basis for all smooth and time
dependent source free solutions. These proofs promote the study
of imaginary sources to a new and crucial viewpoint for
understanding time dependent backgrounds. From our field theory
examples we further propose a completeness conjecture that every
regular solution to a field equation of motion has a
corresponding imaginary source configuration.  We define charges
for spacelike sources and show their compatibility with the usual
charge definition for timelike sources.  Many new non-singular
time dependent field theory solutions are discussed, including
Wick rotations of abelian instantons which demonstrate a close
relationship between electric-magnetic duals and analytic
continuation.}

\preprint{
{\normalsize{\tt hep-th/0510217}}\\
{\normalsize UT-Komaba/05-12}\\
{\normalsize DAMTP-2005-102}\\
{\normalsize NSF-KITP-05-86}
}

\begin{document}

\section{Imaginary Sources}

\subsection{General concept}

Time-dependent backgrounds are a big issue for reconciling string
theory with cosmology, which has been a driving force for their
study in recent years. A time-dependent boundary state Sen
described as rolling tachyon \cite{roll} was obtained by a Wick
rotation, and since boundary states generically act as sources for
closed strings including gravity, this necessarily introduces
\cite{sthermo,imagdbrane, greg1, cards1, twisted} a notion of {\it
sources at imaginary spacetime}, which we call {\it imaginary
sources} in short. This concept has been used in various
situations implicitly, for example in Wick rotation of gravity
solutions, etc. In this paper, we study how the concept of
imaginary sources is general and fundamental.

Inserting sources in imaginary spacetime in various field theories
create many time-dependent classical solutions. While discovering
new solutions is interesting, the question is how much about time
dependent backgrounds do we expect to learn from their study. As
a general remark, first of all, the fact that these solutions do
have sources under analytic continuation is already fascinating,
and more importantly generalizes a host of previous sources such
as black holes, electrons, branes, strings to imaginary cases.
Because of the plethora of discussions of these objects, the fact
that their imaginary versions are consistent should give rise to
much interesting study.

A priori time-dependent solutions obtained by the analytic
continuation of sources seem to some degree artificial. However,
we claim that it is opposite: sources in imaginary spacetimes have
full flexibility in reproducing all the regular solutions to
homogeneous equations of motion, and will give a proof of this
{\it completeness conjecture} for several field theories.

Once assuming that this conjecture is true for all field theories,
one important question is how we can recover the information of the
imaginary sources from given time-dependent solutions. A step toward
answering this question will be given as a new definition of charges for
time-dependent configurations.

\vspace{5mm}
\noindent
\underline{Sources at imaginary spacetime}

Let us first review the basic properties of analytic continuation
and introduce the concept of imaginary sources. To describe a physical
system we take as our starting point a set of real differential
equations with arbitrary number of derivatives with respect to
$\{x_i\}$ for $i=1,...,n$.  The solutions of these equations are
real distributions $\Phi$ and schematically we write the situation
as
\begin{equation}
D_R \Phi_R=0 \ .
\end{equation}
This solution can be formally promoted to the entire complex plane
by changing all instances of real coordinates $\{x_i\}$ by the
complex extension $\{z_i\}$ in both the differential equation and
the solution
\begin{equation}
D_C \Phi_C =0 \ .
\end{equation}
It is then possible to restrict the differential equation to a
subset of the complex plane of real dimensionality $n$.  Generally
this projection produces complex differential equations but if this
is done carefully then this restriction will generate a new real
solution
\begin{equation}
D_P \Phi_P =0 \ .
\end{equation}
The statement of analytic continuation can then be rephrased as
follows.  Take a real differential equation, complete it in the
analytic plane and then find a real restriction of this
complexified system so ``Analytic continuation$=$Completion to
analytic plane $+$ projection to real subset''.
\begin{eqnarray*}
&\Phi_C& \\
\textup{complexify} \;\;\nearrow && \searrow \;\;\textup{projection}\\
 \Phi_R\ \hspace{3mm}   &-\!\!\!-\!\!\!-\!\!\!\longrightarrow&
  \hspace{5mm} \Phi_P\\
 &analytic&\\[-2mm]
 &\!\!\!\!\!\! continuation\!\!\!\!\!\!&
\end{eqnarray*}

The general statement of our paper is that an imaginary source can
be regarded as follows.  A solution of the differential equation
with imaginary source
\begin{equation}
D(ix_1, x_2,...) \Phi(ix_1,x_2,...)= \textup{Imaginary Source S
localized at }(ix_1=s, x_2=0,...)
\nonumber
\end{equation}
is strictly speaking a solution without source for real values of
$x_1$.  For real values of $x_1$ the above source $S=0$. However
the (inverse) analytic continuation of this solution does have a
source for real values
\begin{equation}
D(x_1,x_2,...)\Phi(x_1,x_2,...)= \textup{Source S localized at }
(x_1=s,x_2=0,...) \ .
\nonumber
\end{equation}
Note that we do not insist that these imaginary source solutions
come from Wick rotations of the same Euclidean or Lorentzian
metrics.  Following our procedure, which is also discussed in
Ref.~\cite{sthermo}, we shall also find a way to implicitly define an
analytic continuation of the delta function which we otherwise would
have difficulty in handling.

This procedure of using imaginary sources will be very useful in
generating non-singular solutions to technically source-less
differential equation.  In particular these solutions will
generically be time dependent. This comes from the simple
observation that if a source is placed at a complex value of time,
then there is a notion of approaching the source and receding from
the source during the evolution of real time.

\vspace{5mm}
\noindent
\underline{Completeness of the imaginary sources}

In this paper we also discover the important fact that in several
classes of theories, imaginary sources serve as a complete basis
for all regular and time dependent solutions!  Analysis of
imaginary sources therefore forms a new framework for
understanding time dependence.  For linear systems, one may
heuristically argue for the validity of this statement as
follows. We know that solving the homogeneous equations of motion
with inhomogeniety given by arbitrary source, $S(x)$, can be
reduced to finding the Green's function solution, $G(x,x_0)$. The
general solution is then written as an integral of the source
density times the Green's function. However if the source is zero
over the real coordinate values then apparently this method
fails.  Yet this is not the case.  Although at first sight it
appears that Green's function techniques are not applicable in
solving the homogeneous equations of motion, on the contrary we
will argue that in fact Green's functions can be used to solve
for all solutions.

The point is to properly understand the statement that sources which
are zero valued over the real coordinate values are not necessarily
zero over the complex values.  When we write down a source $S(x)$ it
can be written as an integral over delta functions
\begin{equation}
S(x)=\int d^n x_0 \ S(x_0) \delta(x,x_0) \ .
\end{equation}
However we can examine sources which are zero over the reals but
with support over the complex extension of the coordinates.  In this
case we have the identity
\begin{equation}
0=\int d^n x_0 S(x_0) \ \delta(C(x,x_0))
\end{equation}
when $C(x,x_0)$ is a function with no real roots.  This method
clearly creates all possible real sources and so one may reason that
with the proper definition of analytic continuation, this method
exhausts all possible ways of writing zero for real coordinate
values.

It is the goal of this paper to further describe these imaginary
source techniques and to show the statement above can be made
rigorous at least in certain cases.  Further, this raises the
possibility of a completeness conjecture for other theories
including non-Abelian gauge theories, gravity and string theory.
We will conjecture, based on our evidence, that {\it for every
smooth solution to the inhomogeneous equations of motion there is
a distribution of imaginary sources}.  This statement, when true,
promotes the study of imaginary sources to a crucial ingredient of
all general space and time dependent backgrounds.

\vspace{5mm}
\noindent
\underline{Charges of time-dependent configurations}

Once various time-dependent solutions are obtained, a natural
question is how we can characterize those configurations. In
static cases, one can specify the solutions by local charges. In
our strategy, solutions are obtained by Wick rotation, thus it is
plausible that some quantity like the charge can be defined also
by the Wick rotation. This should be a new notion because it is
not conserved in the usual sense but defined in the full
spacetime. Since this new definition of ``S-charge'' is obtained
through the same type of Wick rotation used for obtaining the
time-dependent solutions, it certainly captures the sources in the
imaginary spacetime, as we will see. This new notion may be
extended to any theory including gravity, thus possibly providing
a new concept of ``ADM S-mass'' for time-dependent gravity
backgrounds. There are other definition of the S-charges in the
literature \cite{stro, greg1, cards1}, and we will compare them
with our definition.

The organization of this paper is as follows. In the next
subsection we start with a simple solution with sources in
imaginary time as a demonstration and to serve as a basis for the
following sections. In Sec.~\ref{sec:completeness}, we state our
completeness conjecture on imaginary sources and give a proof for
Klein-Gordon theory and Maxwell electrodynamics in any
dimensions. In Sec.~\ref{sec:charges}, a new definition of
``charge'' capturing properties of time-dependent backgrounds is
given, and its possibilities are explored. In
Sec.~\ref{sec:more}, we present a number of time-dependent
solutions generated by the imaginary sources, in the Klein-Gordon
systems and the Maxwell electrodynamics, including Wick rotation of
electrons, monopoles and instantons.

\subsection{A free field theory example}\label{sec:freeex}

To illustrate the general concept of the imaginary source, in this
subsection we will study imaginary sources in a simple but important
example. We consider the massless real Klein-Gordon system in $n$
dimensions. Although this example seems too simple, we remark that
this often arises as the first order approximation to a variety of
systems from gravity to condensed matter to string theory.
 Interesting other examples will be given in Sec.~\ref{sec:more},
including electromagnetic wave solutions of various shapes.

The Laplace equation of motion is
\begin{eqnarray}
\sum_{i=1}^{n} \left(\frac{\p}{\p x_i}\right)^2 \phi(x_1, x_2,
\cdots, x_n) =0.
\label{laplacendim}
\end{eqnarray}
We complexify the coordinate $x_n$ into $z_n \equiv x_n + i x_0$,
where $x_0$ is introduced as an imaginary part of $z_n$. If $\phi(x_i)$
is a solution to the above Euclidean equation (except at possible
singularities corresponding to sources), then simply replacing $x_n$
by $z_n$ in the solution as
\begin{eqnarray}
 \phi(x_1, \cdots, x_{n-1}, z_n)
\label{comsol}
\end{eqnarray}
gives a solution of the holomorphic equation
\begin{eqnarray}
\left[ \left(\frac{\p}{\p z_n}\right)^2 + \sum_{i=1}^{n-1}
\left(\frac{\p}{\p x_i}\right)^2 \right] \phi(x_1, \cdots, x_{n-1}, z_n)
=0.
\end{eqnarray}
Then, extracting just the imaginary part of $z_n$ we get the
``Wick rotated'' solution of the Lorentzian wave equation of motion,
\begin{eqnarray}
\left[ -\left(\frac{\p}{\p x_0}\right)^2
+\sum_{i=1}^{n-1} \left(\frac{\p}{\p x_i}\right)^2
\right]
\phi(x_1,\cdots, x_{n-1}, ix_0)=0.
\end{eqnarray}
In this subsection, we see explicit examples of this procedure in 1+3
dimensions and find
interesting time-dependent solutions which are spherical waves.

The first example is a Wick rotation of the simplest and the most
familiar solution to the Klein-Gordon equations.  To obtain a
time-dependent solution in 1+3 dimensions, we start in 4
dimensional Euclidean space.  In this case we can solve the four
dimensional Laplace equation (\ref{laplacendim}) by
\begin{eqnarray}
 \phi = \sum_{a=1}^{N} \frac{C_a}{x_1^2+x_2^2 + x_3^2 + (x_4-s_a)^2 }.
\label{orisol}
\end{eqnarray}
This is the harmonic function with $N$ sources located at
$x^{I}=(0,0,0,s_a)$. The charge of each source is given by the
constant $C_a$. Starting from this solution, we immediately obtain a
time-dependent source free solution in 1+3 dimensions to the
Klein-Gordon system by applying our above procedures. The complex
extension of the configuration (\ref{orisol}) is
\begin{eqnarray}
 \phi = \sum_{a=1}^{N} \frac{C_a}{x_1^2 + x_2^2 + x_3^2 + (z_4-s_a)^2 }.
\label{cpxsol}
\end{eqnarray}
And thus we obtain a complex valued solution for the Klein-Gordon (KG)
system as
\begin{eqnarray}
  \phi = \sum_{a=1}^{N} \frac{C_a}{(ix_0-s_a)^2 + x_1^2 + x_2^2 + x_3^2}.
\end{eqnarray}

Since this analytic continuation in general does not give a real
valued solution, our procedure does not seem to make sense
classically. As we will show below, however, the complex
conjugate of the above Wick rotated solution is also a solution.
So the real part of the above solution $\phi$ also solves the
Klein-Gordon equation thus giving the real valued solution we are
looking for. Physically, we are stating that although Wick
rotating one source gives an unphysical complex valued solution,
the Wick rotating of {\it{pairs}} of sources into imaginary
coordinates can give a real solution.  Adding such solutions
together is possible due to the linearity of the Klein-Gordon
equations.  Even in a nonlinear theory such as string theory,
pairs of imaginary D-branes were for example discussed in
Ref.~\cite{sthermo, imagdbrane, greg1} and odd numbers of sources
were discussed in Ref.~\cite{cards1}.

If we look at the above general complexification procedure, then we
can understand that we may complexify $x_4$ by replacing it by
$\overline{z}_4 = x_4 - ix_0$, not by $z_4$. Even with this
replacement, the resultant KG operator is the same. Thus, from this
procedure we obtain a new solution to the KG equation
\begin{eqnarray}
  \widetilde{\phi} =
\sum_{a=1}^{N} \frac{C_a}{(-ix_0-s_a)^2 + x_1^2 + x_2^2 + x_3^2}.
\end{eqnarray}
If the parameters $C_a, s_a$ are real, then this is just the
complex conjugate of the previous solution. Thus,
$\widetilde{\phi} = \phi^*$
and the real solution is $\phi+\phi^*$.
It is also possible to take the imaginary part and multiply it by $i$
to get an alternate real solution, $-i(\phi-\phi^*)$. In other words,
it is also possible to take the $C_a$'s to be imaginary.

To be more concrete, let us consider the simplest case in which
$C_1 = C_2 = C/2$ and $s_1 = -s_2 = s$ which are real; this is the
same as using our analytic continuation to find a complex solution
and then adding to it its complex conjugate. (It seems that, after
the rotation procedure, we put sources at imaginary locations
$(x_0,x_1,x_2,x_3)=(\pm is, 0,0,0)$.  This is in fact true
although it is not naively so.  We come to this point later.) The
solution is
\begin{eqnarray}
 \phi &=& \frac{C/2}{(ix_0-s)^2 + x_1^2 + x_2^2 + x_3^2}
+\frac{C/2}{(ix_0+s)^2 + x_1^2 + x_2^2 + x_3^2}
\nonumber \\
  &=& \frac{C(-x_0^2 + s^2 + x_1^2 + x_2^2 + x_3^2)}
{(-x_0^2 + s^2 + x_1^2 + x_2^2 + x_3^2)^2 + 4 s^2x_0^2 }.
\label{sol1phi}
\end{eqnarray}
This solution has both positive and negative valued regions. For
real values of $C$ the second term is just the complex conjugate of
the first term, thus one may understand that this procedure is
simply adding the complex conjugate to make the entire solution real
valued. Again, more physically stated, this is the same as Wick
rotating a pair of similarly charged sources.

Let us consider the singularity structure of this solution. To begin
with we look at the locus where $\phi$ diverges.  This only occurs when
the denominator is zero and more precisely only when the two squared
terms are each separately zero
\begin{eqnarray}
-x_0^2 + s^2 + x_1^2 + x_2^2 + x_3^2= 4 s^2x_0^2 =0.
\end{eqnarray}
It is easy to see that when $s\neq 0$ this has no real solution.
Therefore we conclude that the solution is regular over all real
coordinate values.\footnote{Note however that this expectation of
regular sources is not always correct.  As an instructive example,
let us consider a source at a imaginary location in the Euclidean 4
dimensional Klein-Gordon system,
\begin{eqnarray}
 \phi = \frac{C}{(x_1)^2+(x_2)^2+(x_3)^2+(x_4-is)^2 } + \mbox{c.c.}
\end{eqnarray}
Then one finds that the singularity exists at
\begin{eqnarray}
 x_4=0, \quad \mbox{and} \quad
(x_1)^2+(x_2)^2+(x_3)^2 = s^2.
\end{eqnarray}
So this supposedly alternative method of placing sources at
imaginary points has real singularities and does not give rise to
regular solutions. In fact this solution is very unusual in that
asymptotically it can be made to carry the same charge as a usual
static source.  One might wonder if this kind of static solution
has any relationship to spinning black holes with ring shaped
singularities.}

\begin{figure}[t]
\begin{center}
\begin{minipage}{7cm}
\begin{center}
\includegraphics[width=7cm]{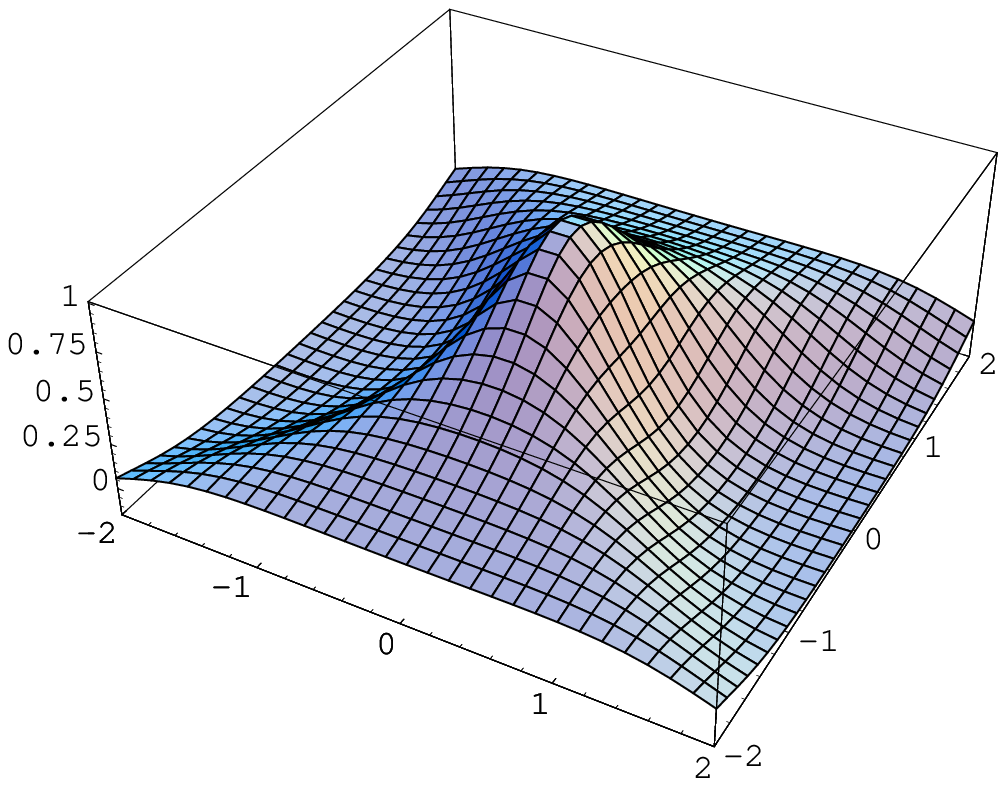}
\caption{A 1+1 dimensional slice of the solution for two sources with the
same charge is smooth and shows a 2-sphere shrinking and expanding.}
\label{fig1}
\end{center}
\end{minipage}
\hspace{5mm}
\begin{minipage}{7cm}
\begin{center}
\includegraphics[width=7cm]{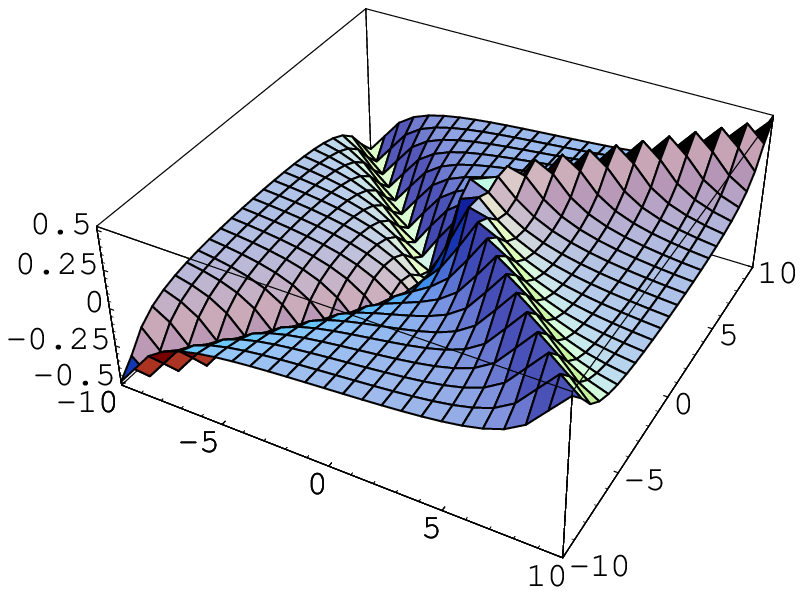}
\caption{For two opposite charges, wave flips over in time which
runs left to right.} \label{flip-wave-far}
\end{center}
\end{minipage}
\end{center}
\end{figure}

We have plotted the function $\phi$ in terms of $x_0$ and $x_1$ while
putting $x_2 = x_3=0$, in Fig.~\ref{fig1}. It is in fact regular,
and one can see how traveling waves come in, collide and then go
out with most of the activity centered near the light cone. This
is a non-singular collision of two wave packets traveling in
opposite directions.  But we are in 1+3 dimensions, so more
precisely speaking, this is a formation of a lump by
time-dependent shrinking 2-sphere. It shrinks to form a lump at
$x_0=0$ and then decays to a spreading out 2-sphere.

Let us study the sources in this Wick rotation procedure. When we
complexify the solution as (\ref{cpxsol}), the locations of the
sources are in fact also extended in the complex plane. For our
present example with $C_1 = C_2 = C/2$ and $s_1 = -s_2 = s$, the
location of the singularities are simply obtained by examining where
the denominator of the complex solution vanishes,
\begin{eqnarray}
 (ix_0\pm s)^2 + x_1^2 + x_2^2 + x_3^2 =0.
\end{eqnarray}
The location of the singularities in this KG system are at
\begin{eqnarray}
 x_0 = \sqrt{x_1^2 + x_2^2 + x_3^2} \pm i s
\end{eqnarray}
Here we make a distinction between singularities and sources. The
sources are point-like and at $x_0=\pm is$, they are distributed in
imaginary space in terms of the $(x_0,x_1,x_2,x_3)$ coordinates.
These sources generate the rest of the light-cone type
singularities.
Since the singularities are all at complex coordinates values, we
can thus describe this as saying there are no real sources in the
Klein-Gordon system. This is the reason why we obtained a
source-free time-dependent solution of the Klein-Gordon system.

Having only dealt with the real part of the complexified solution,
clearly there is also another solution which we can generate from
the analytic continuation procedure, which is to take the case of
two opposite charges with $C_1=-C_2=C/2i$ for real $C$
\begin{eqnarray}
 \phi &=& \frac{C/2i}{x_1^2 + x_2^2 + x_3^2+(ix_0-s)^2}
-\frac{C/2i}{x_1^2 + x_2^2 + x_3^2+(ix_0+s)^2}
\nonumber \\
  &=& \frac{C s x_0}
{(-x_0^2 + s^2 + x_1^2 + x_2^2 + x_3^2)^2 + 4 s^2x_0^2 }.
\label{sol2phi}
\end{eqnarray}
This solution is non-singular with the same ``imaginary''
singularity type
structure as the previous case but just with different charges. A
plot of the solution is shown in
Fig.~\ref{flip-wave-far}.

Note that the solution does not form a lump at the origin, is zero
for $x_0=0$ and has a maximum and a minimum.  This solution can be
understood as the scalar field theory analogue of the gravity
solutions generated from diholes
of Ref.~\cite{cards1, greg1}.  Apparently the maximum and minimum
values of the scalar field here correspond to the creation of large
curvatures and therefore the horizons in the gravity solutions. The
field theory solutions are different however in that they do not
have an interpretation as the creation of a lower dimensional object
as in the gravity case.

Of course it is possible that other solutions will generate divergences
on the real $x_0$ axis even after analytic continuation.  In
such a case the Wick rotated system will also have a real source.
For example, we can take the displacement of the sources in
imaginary time to zero $s\rightarrow 0$, in which case these
non-singular time dependent solutions to the wave equation do have a
singular limit. These are the scalar field counterparts to the
singular S-branes found in Ref.~\cite{stro} where there is a spacelike
source existing at one moment in time.  We will further discuss the
usefulness of imaginary sources in the following sections and define
a new way to calculate their charges.

%%%%%%%%%%%%%%%%%%%%%%%%%%%%%%%%%%%%%%%%%%%%%%%%%%%%%%%%%%%%%%%%%%%%%%
%%%%%%%%%%%%%%%%%%%%%%%%%%%%%%%%%%%%%%%%%%%%%%%%%%%%%%%%%%%%%%%%%%%%%%
%%%%%%%%%%%%%%%%%%%%%%%%%%%%%%%%%%%%%%%%%%%%%%%%%%%%%%%%%%%%%%%%%%%%%%
%%%%%%%%%%%%%%%%%%%%%%%%%%%%%%%%%%%%%%%%%%%%%%%%%%%%%%%%%%%%%%%%%%%%%%

\section{Imaginary Source Completeness}
\label{sec:completeness}

\subsection{Our conjecture}

We have seen that interesting time-dependent source-free solutions
to the massless KG equation of motion can be obtained by applying
the imaginary source technique. Then the following question
naturally arises: what kind of solutions can we obtain from the
technique? How general are the imaginary source solutions? Our
answer to this question is the following conjecture:

{\it Any regular solution to field equations in any theory has
a corresponding distribution of sources in imaginary spacetime.}

This statement is quite nontrivial. In this section, we provide a
proof of this conjecture for wide variety of linear systems ---
massless/massive Klein-Gordon theories, and the Maxwell theory, in
any dimensions. It will be shown that for every smooth solution of
these theories, there is a corresponding imaginary source
configuration.  Surprisingly we will find that pairs of
oppositely charged sources generate all the time dependent
solutions in these cases, including the like charged source
solutions.  Hence dipole configurations play the crucial role as
the basic ingredients of imaginary sources as opposed to single
charges. With these insights we see that the status of imaginary
sources is therefore promoted from curiosity to a powerful new
viewpoint.

\subsection{Proof for 1+0 dimensional Klein-Gordon system}
\label{sec1+0}

A situation where the proof of the completeness conjecture can be
explicitly and easily provided is the lowest dimensional example,
that is, $1+0$ dimension. First let us consider the massless KG
system.

The Euclidean solution with sources at $x=\pm s$ is easily written
as
\begin{eqnarray}
 \phi= \left\{
\begin{array}{ll}
-iA x & (-s<x<s) \\
iAs & (x<-s) \\
-iAs & (s<x)
\end{array}
\right.
\end{eqnarray}
The magnitude of the source is given just by $\pm iA$. For
simplicity we take the limit $s\to\infty$ to get just $\phi=-iAx$. The
Wick rotation $x \to it$ gives
\begin{eqnarray}
 \phi= A t
\end{eqnarray}
which is in fact a generic solution to the equation of motion in 1+0
dimensional KG system once a shift in real $t$ is included.

The same argument holds for massive case. We put sources at $x=\pm
s$ in Euclidean space with opposite charges, and then take the
$s\to\infty$ limit to get\footnote{The magnitude of the
corresponding sources depends on what kind of asymptotic condition
we chose. We may require that asymptotically the solution is zero
(approaching zero exponentially), but this doesn't fix the
magnitude of the charges completely. There remains arbitrariness
of the choice of the charge amount, but anyway we are taking
$s\to\infty$ limit and so this is not relevant to the following
argument.}
\begin{eqnarray}
 \phi = A e^{mx} + Be^{-mx} \ .
\end{eqnarray}
Then the Wick rotation gives
\begin{eqnarray}
 \phi = A e^{imt} + Be^{-imt}
\end{eqnarray}
which is in fact a generic solution to the massive KG equation of
motion.

After giving a proof for $1+1$ dimensions in the next subsection,
in Sec.~\ref{sec:proofany} we will prove the completeness of the
massless/massive KG theories in any dimensions.  Although the
above proof in this subsection is straightforward, it will in
fact also prove very useful.  Before proceeding for that, here we
present an issue of Wick rotation. Note that in the above we took
$s\to\infty$ limit, but let us consider what happens if we didn't
take this limit. It seems that this limit is unnecessary because
the change of $s$ does not modify the solution around the origin,
$\phi= -iA x$. After the Wick rotation we look only at the
imaginary axis of $x$. However, this sloppy argument contains a
subtle issue which will be explained in Sec.~\ref{infinitearraysec}.
We have to take the $s\to\infty$ limit for completeness.

\subsection{Proof for 1+1 dimensional massless Klein-Gordon system}
\label{secplane}

For the massless KG system in 1+1 dimensions, the most general
solutions to the equation of motion are known,
\begin{eqnarray}
\phi = f(x-t) + g(x+t)
\label{genericsol}
\end{eqnarray}
where $f$ and $g$ are arbitrary functions. We show that for any
functions, $f$ and $g$, we have a corresponding distribution of
sources in imaginary spacetime.

Our strategy in Sec.~\ref{sec:freeex} for dealing with imaginary
sources was to start with a Euclidean version of the field theory
and to consider harmonic functions in that Euclideanized space. In
two Euclidean dimensions spanned by $x$ and $y$, the harmonic
function is $\phi=\log(x^2+y^2)$ and so an array of such charges
separated in the $y$-direction produces the field
\begin{eqnarray}
\phi=\sum_n \frac{q_n}{4\pi} \ln(x^2+(y-y_n)^2).
\end{eqnarray}
Here the source term should be given consistently by
\begin{eqnarray}
\left(
\p_x^2 + \p_y^2
\right) \phi = \sum_n q_n \delta(x)\delta(y-y_n).
\end{eqnarray}
Such a potential can also be interpreted as being generated by an
infinitely long linear static source in 1+3 dimensions. Taking
the analytic continuation $y\rightarrow -it$ of such a
configuration we get
\begin{equation}
\phi=\sum_n \frac{q_n}{4\pi} \ln(-(t-iy_n)^2+x^2) \ ,
\end{equation}
for which the source is given by
\begin{eqnarray}
 \sum_n q_n \delta(y)\delta(-it-y_n).
\end{eqnarray}
Evidently this means that the source is located at imaginary time.

\begin{figure}[t]
\begin{center}
\begin{minipage}{7cm}
\begin{center}
\includegraphics[width=7cm]{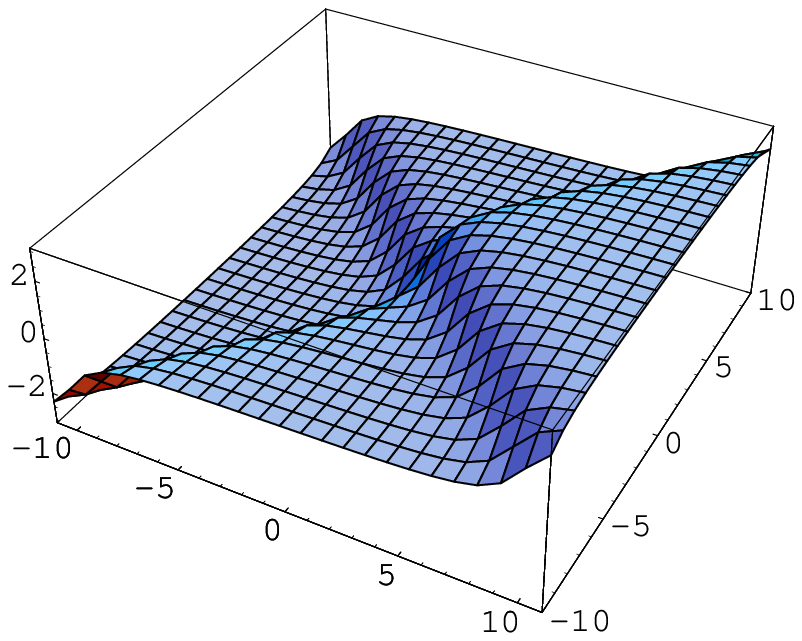}
\caption{
 Imaginary part of log wave with two oppositely charged imaginary
sources at $t=\pm i$ is asymptotically flat at infinity.  It is an
upside down wave coming in from past infinity and then it flips to
an upright wave; time runs left to right in the figure.}
\label{fig4}
\end{center}
\end{minipage}
\hspace{5mm}
\begin{minipage}{7cm}
\begin{center}
\includegraphics[width=7cm]{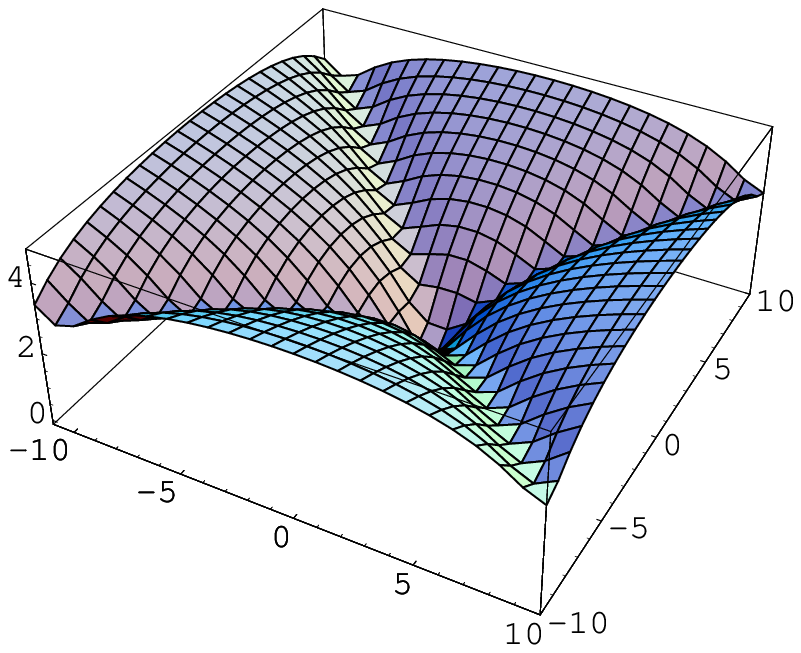}
\caption{Real part of log wave with two same charged imaginary
sources at $t=\pm i$; time runs left to right in the figure.}
\label{fig5}
\end{center}
\end{minipage}
\end{center}
\end{figure}

For illustration, begin with the case of two sources separated by
a distance $s$
\begin{equation}
\phi=\frac{q}{2\pi i}\left[
\ln(x^2+(y-s/2)^2) - \ln (x^2+(y+s/2)^2)
\right].
\end{equation}
The normalization of the charge has been chosen for the later purpose.
The Wick rotation gives a Lorentzian solution which is completely
regular,
\begin{equation}
\phi=\frac{q}{2\pi i}\left[
\ln(-(t-is/2)^2+x^2) - \ln (-(t+is/2)^2+x^2)
\right]
\end{equation}
which can be simplified to
\begin{equation}
\phi=\frac{q}{\pi} \arctan\frac{ts}{x^2-t^2+s^2/4} \ .\label{lines-1}
\end{equation}
This solution, as shown in Fig.~\ref{fig4}, is related to two
opposite charged sources and is asymptotically flat.  Now because
the arctangent function is multi-valued, we have to specify the
period of the values which the function takes.  For the solution
to be smooth also in derivatives we choose the branches of the
arctangent to take
\begin{eqnarray}
\left\{
\begin{array}{ll}
0<\phi/q \leq 1 & (t>0) \\
-1 \leq \phi/q <0& (t<0) \ .
\end{array}
\right.
\end{eqnarray}
Choosing branches in this fashion amounts to three different
branches of arctangent which smoothly fit together. The solution
is created to be spatially symmetric under $x\rightarrow -x$.
However it is clearly also antisymmetric in time $t\rightarrow
-t$ and therefore imaginary time so the solution must vanish at
$t=0$.

We could also take the two like charged solutions instead,
whose Wick rotation would give, up to a proportional constant,
\begin{equation}
\phi=\ln(\sqrt{(x^2-t^2+s^2/4)^2+t^2s^2}) \ .
\end{equation}
Although this solution will not turn out to play an important role
in the following discussion, this solution is plotted in
Fig.~\ref{fig5} and is new in that the potential is not flat at
infinity.

Let us look at the asymptotic region of the solution
(\ref{lines-1}) in more detail.  We can take a limit
$x+t\to\infty$ while keeping $x-t$ finite and fixed, which gives
\begin{eqnarray}
 \phi \to \frac{q}{\pi} \arctan\frac{s/2}{x-t}.
\label{asymplines}
\end{eqnarray}
This shows that the asymptotic region is in fact a wave packet
whose width is given by $s$ and whose amplitude is not decaying
but a constant. In other words, the asymptotic solution is in
fact a (localized) plane wave. Because this plane wave has an
arbitrary real parameter $s$ which is the width, we can
generically expect that this S-brane solution may form a basis
for a function space in 1+1 dimensions, as the plane waves do.
Below we present a precise proof of the completeness.

\begin{figure}[t]
\begin{center}
\begin{minipage}{13cm}
\begin{center}
\includegraphics[width=13cm]{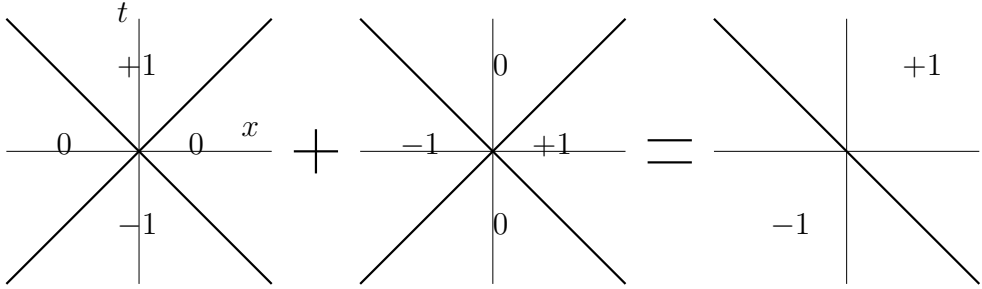}
\put(-327,100){$t$}
\put(-280,57){$x$}
\put(-327,80){$+1$}
\put(-327,20){$-1$}
\put(-300,50){0}
\put(-350,50){0}
\put(-185,80){0}
\put(-185,20){0}
\put(-170,50){$+1$}
\put(-220,50){$-1$}
\put(-80,20){$-1$}
\put(-30,80){$+1$}
\caption{Addition of the original solution with imaginary time source
 and the new solution with imaginary space source gives a purely
 left-moving wave.} \label{fig-ty}
\end{center}
\end{minipage}
\end{center}
\end{figure}

As suggested in the generic form of the solution (\ref{genericsol}),
one can show that the solution (\ref{lines-1}) is written
as a collision of two wave packets,
\begin{eqnarray}
 \phi = \frac{q}{\pi}
\left(
\arctan\frac{-s/2}{x+t}
-\arctan\frac{-s/2}{x-t}
\right)
\label{sbranedec}
\end{eqnarray}
not asymptotically but exactly. Here the arctan functions are chosen
to take values in $0<\arctan<\pi$. Let us take the thin wall limit
$s\to 0$, then the solution becomes
\begin{eqnarray}
 \phi = q \left(\theta(x+t)-\theta(x-t)\right).
\label{leftright}
\end{eqnarray}
The corresponding sources are distributed at $(t,x)=(\pm s, 0)$,
or more precisely, the source is given by the distribution
function
\begin{eqnarray}
\lim_{s\to 0} \frac{2q}{i}\delta(x)\left[
\delta(-it-s/2)-\delta(-it+s/2) \right].
\end{eqnarray}
The solution in this limit(\ref{leftright}) seems to serve as a
basis of the generic solution (\ref{genericsol}), but
unfortunately it leaves a constraint between the amplitudes of
the left and right moving modes.  This difficulty can not be
overcome by adding combinations of the same sources displaced
relative to each other since each source suffers from the same
left-right mode symmetry.  To overcome this difficulty, we must
introduce a new solution whose imaginary source is located at
$(t,x)=(0,\pm is)$, that is, the solution obtained by exchange of
the roles of $t$ and $x$.  What we get as a result of the sum of
those four sources is the solution (see Fig.~\ref{fig-ty})
\begin{eqnarray}
 \phi = q (2\theta(x+t)-1).
\label{leftmover}
\end{eqnarray}
This is a purely left-going wave of a step function.
Interestingly, this procedure to get a purely left-mover
introduces the new idea of sources at {\it imaginary space}.

At this stage it is quite easy to see that this forms a basis for
generating all the solutions (\ref{genericsol}), because a derivative of
this step function
is in fact a delta function $\delta(x+t)$. Thus, adding various
imaginary source distributed as the following
source term
\begin{eqnarray}
&&\lim_{s\to 0}\int \! dx_0dt_0\; \left[ \frac{2q(x_0,t_0)}{i}
\delta(x-x_0) \biggl( \delta(-i(t-t_0)-s/2)-\delta(-i(t-t_0)+s/2)
\biggr) \right.
\nonumber \\
&& \hspace{20mm}\left.
+\frac{2\tilde{q}(x_0,t_0)}{i} \delta(t-t_0)
\biggl(
\delta(-i(x-x_0)-s/2)-\delta(-i(x-x_0)+s/2)
\biggr)
\right]\hspace{10mm}
\end{eqnarray}
we may reproduce the generic solution (\ref{genericsol}). Here the
second term is given by the exchange of the roles of $t$ and $x$,
and we let the source amplitude $q$ and $\tilde{q}$ to be dependent
also on
the real spacetime $(x_0,t_0)$.
In fact, if we choose
\begin{eqnarray}
&& q(x_0, t_0) = -f'(x_0-t_0)\delta(x_0 + t_0)+
g'(x_0+t_0)\delta(x_0 - t_0), \\
&& \tilde{q}(x_0, t_0) = f'(x_0-t_0)\delta(x_0 + t_0)+
g'(x_0+t_0)\delta(x_0 - t_0),
\end{eqnarray}
a straightforward calculation reproduces (\ref{genericsol}). This is
a proof of the imaginary source completeness in the 1+1 dimensional
massless Klein-Gordon system. We see in this non-trivial example the
power and relevance of imaginary sources as being a possible way to
understand all features of time dependence.

\subsection{Proof in any dimensions for Klein-Gordon and Maxwell
  systems}
\label{sec:proofany}

Using the results of the previous subsections, we give a simple
argument to show the completeness of imaginary sources in any
dimensions of the Klein-Gordon system and the Maxwell system.

\vspace{5mm}
\noindent
\underline{Massive Klein-Gordon}

First, let us consider the massive Klein-Gordon system. Any
configuration of the field $\phi$ has a representation in terms of
Fourier basis,
\begin{eqnarray}
 \phi(x)= \int d^{d+1}k \; F(k)e^{ik_\mu x^\mu}.
\end{eqnarray}
For this configuration to be a solution of the equation of motion,
we know that the plane wave is necessarily on-shell,
\begin{eqnarray}
 k_0^2 - k_i^2 - m^2=0 \quad (i=1,2,\cdots,d).
\label{pw}
\end{eqnarray}
Therefore once this Fourier basis is constructed in terms of the
imaginary source in time, a proof of the completeness is given. In
fact it is very easy to obtain the corresponding imaginary source
distribution, if one performs a Lorentz transformation on the
above momentum to get
\begin{eqnarray}
 k_0 = \pm m, \quad k_i=0.
\end{eqnarray}
This plane wave is just the same as what we obtained in the 1+0
dimensional case of subsection~\ref{sec1+0}. So if we distribute
the imaginary source of the previous subsection to also fill all
the real spatial dimensions, and perform the Lorentz
transformation back, then we certainly get the plane wave with
the momentum (\ref{pw}).  Completeness of the imaginary sources
is therefore proven in massive Klein-Gordon theory.

Note that it is not necessary to introduce imaginary space sources
in the massive Klein-Gordon case to remove the left/right moving
symmetry as was needed in the case of massless Klein-Gordon. This
is due to the fact that one can reduce the entire problem of the
massive Klein-Gordon case to 0+1 dimensions which does not have
left/right moving modes.

\vspace{5mm}
\noindent
\underline{Massless Klein-Gordon}

For massless KG theory, we may employ a similar argument. In this case
the on-shell condition for the momentum of the plane wave is
\begin{eqnarray}
 k_0^2 - k_i^2 =0 \quad (i=1,2,\cdots,d),
\end{eqnarray}
and a proper Lorentz transformation leads to
\begin{eqnarray}
 k_0=\pm k_1, \quad k_2 =k_3 = \cdots = k_d=0.
\label{pwm2}
\end{eqnarray}
Having reduced the situation to the case of 1+1 dimensions, we
can apply our completeness proof from Sec.~\ref{secplane}. There
we wrote down explicitly the distribution of imaginary sources
needed to reproduce a plane wave with momentum (\ref{pwm2}).
Applying the inverse Lorentz transformation gives us the desired
source distribution in arbitrary $1+d$ dimensions thereby explicitly
proving the completeness conjecture.

\vspace{5mm}
\noindent
\underline{Maxwell system}

Interesting solutions of the Maxwell system will be given in
Sec.~\ref{secmaxwell}, but here we shall give a proof of the
completeness for this system. The proof follows once we have the
proof for the massless Klein-Gordon system: in a Lorentz gauge
all the equations of motion in the Maxwell system are identical
to that of the massless Klein-Gordon system. This is enough to
know that all the Maxwell regular solutions can have imaginary
source interpretation.  One difference however is that the charge
for Klein-Gordon is a scalar while the charge for Maxwell theory
is a vector/one form.  Lorentz transformations will therefore act
differently on these charges as opposed to the scalar
Klein-Gordon charges.

%%%%%%%%%%%%%%%%%%%%%%%%%%%%%%%%%%%%%%%%%%%%%%%%%%%%%%%%%%%%%%%%%%%%
%%%%%%%%%%%%%%%%%%%%%%%%%%%%%%%%%%%%%%%%%%%%%%%%%%%%%%%%%%%%%%%%%%%%
%%%%%%%%%%%%%%%%%%%%%%%%%%%%%%%%%%%%%%%%%%%%%%%%%%%%%%%%%%%%%%%%%%%%
%%%%%%%%%%%%%%%%%%%%%%%%%%%%%%%%%%%%%%%%%%%%%%%%%%%%%%%%%%%%%%%%%%%%
%%%%%%%%%%%%%%%%%%%%%%%%%%%%%%%%%%%%%%%%%%%%%%%%%%%%%%%%%%%%%%%%%%%%

\section{Capturing Moving Charges}
\label{sec:charges}

In this section we discuss a new definition of S-brane charge
which we apply to our imaginary source solutions and find that it
characterizes the imaginary source solutions. We show that in the
correct limit it reduces to the usual charge definition of a
static charge.  This new definition of S-charge is also compared
with previous definitions \cite{stro, greg1, cards1}.

\subsection{A new definition of S-charge}

\begin{figure}[t]
\begin{center}
\begin{minipage}{7cm}
\begin{center}
\includegraphics[width=7cm]{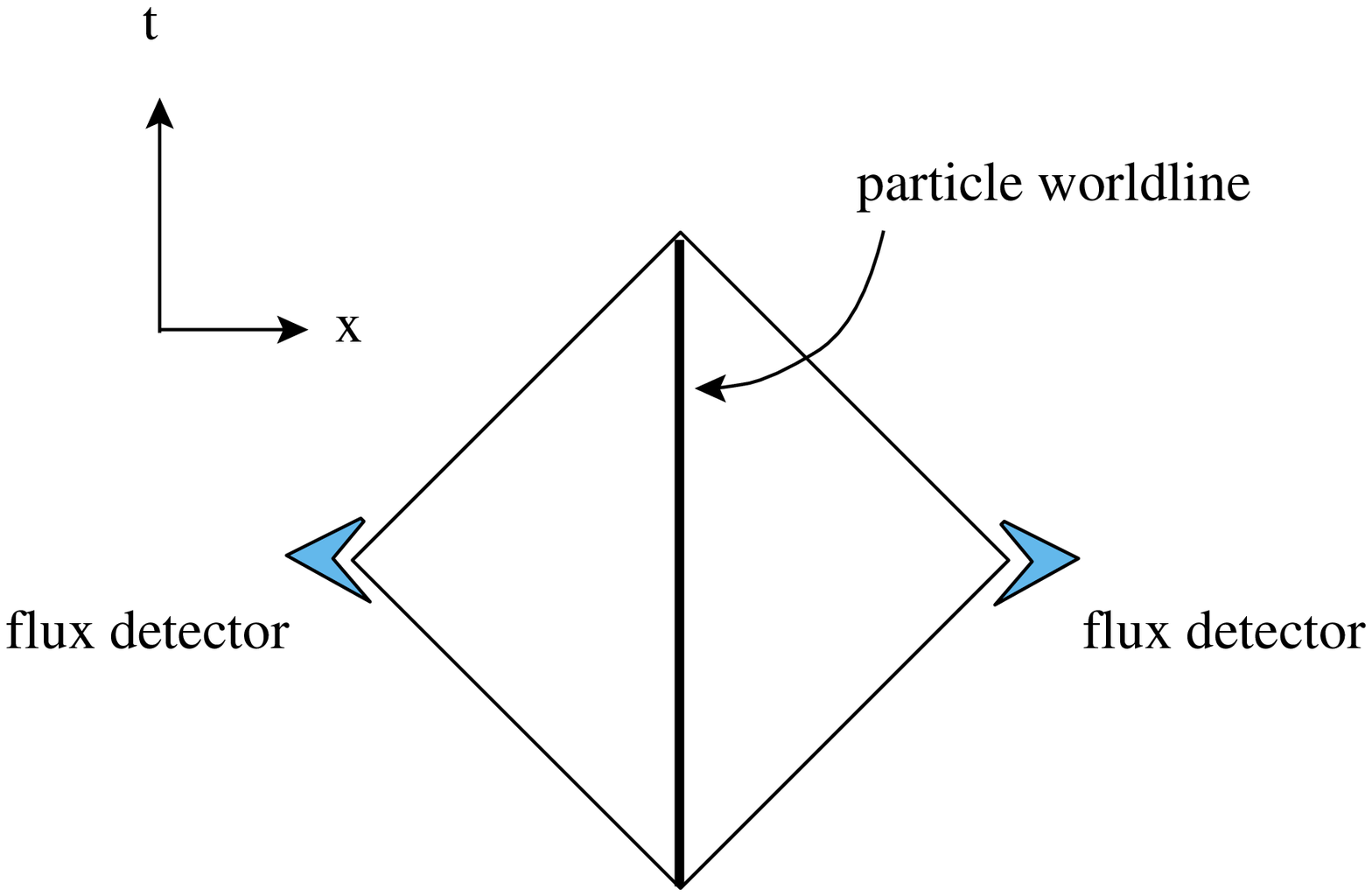}
\caption{In Minkowski space, we can measure the charge of a
stationary particle by surrounding it with flux detectors at spatial
infinity. This is labeled in the Penrose diagram.}
\label{charge-detector}
\end{center}
\end{minipage}
\hspace{2mm}
\begin{minipage}{7cm}
\begin{center}
\includegraphics[width=7cm]{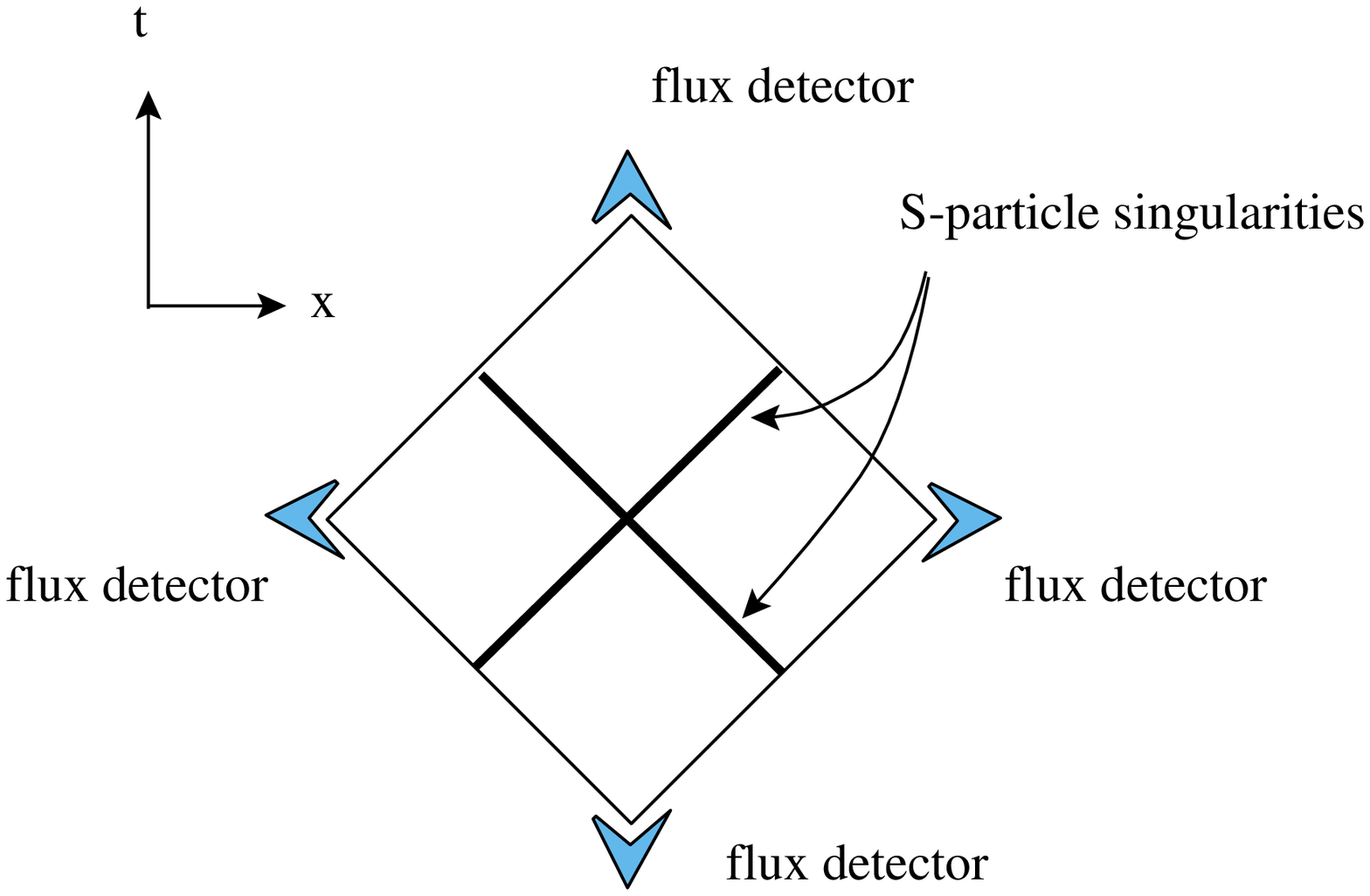}
\caption{To calculate the charge of a spacelike source, we measure
the flux by going to timelike and spatial infinity. The lightcone
singularities generate a source at their intersection point. }
\label{Scharge-detector}
\end{center}
\end{minipage}
\end{center}
\end{figure}

In the case of Euclidean spacetimes, we surround a point by choosing
another distinct point and then applying the rotation group
preserving the distance from the source. This generates a sphere
surrounding the source. One can take the limit where we are far from
the source at spatial infinity and place flux detectors there to
count the charge. This is illustrated in a Penrose diagram in
Fig.~\ref{charge-detector}.

In the case of Lorentzian spacetimes and spacelike sources, the
rotation group is disconnected.  We wish to cover the source with
hyperboloids. In fact this is a generalization of the notion of
surrounding a charge in Lorentzian signature spacetimes. Each
connected piece of the Lorentz group will generate one of three
hyperboloids.

\begin{figure}[t]
\begin{center}
 \begin{minipage}{14cm}
\begin{center}
\includegraphics[width=13cm]{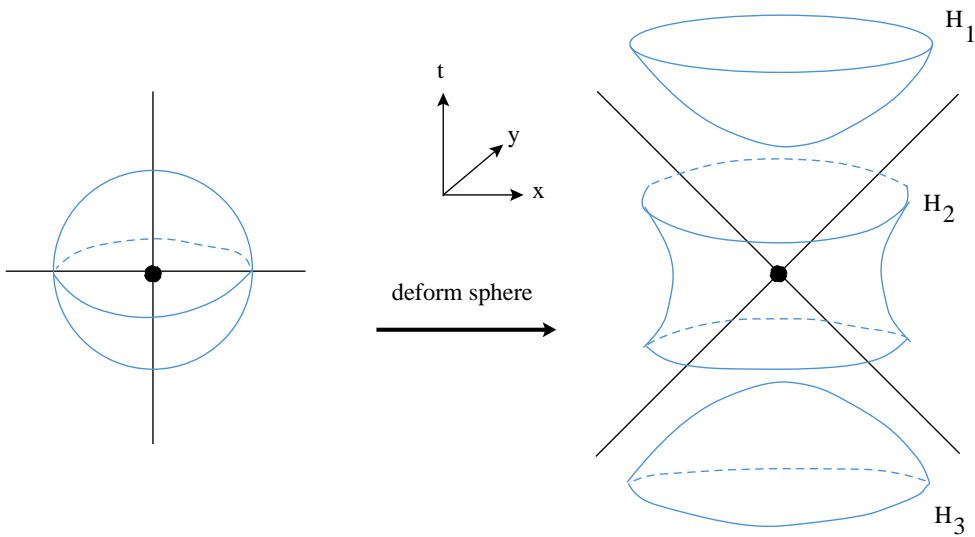}
\caption{To surround a spacelike source we must capture flux in the
time and spacelike directions.} \label{hyperbola-Scharge}
\end{center}
 \end{minipage}
\end{center}
\end{figure}

In 3 dimensions, we have three hyperboloids: $x^2+y^2-\tau^2=a^2$,
which we call $H_2$, is a connected hyperboloid; and
$x^2+y^2-\tau^2=-a^2$ gives two disconnected hyperboloids which
we call $H_1, H_3$ as shown in Fig.~\ref{hyperbola-Scharge}.
If we take the limit of these hyperboloids to infinity, we see
that we should place flux detectors at spatial and timelike infinity
as shown in the Penrose diagram of Fig.~\ref{Scharge-detector} to
completely enclose the source.  Such a notion of surrounding a
spacelike charge makes sense in spacetime as opposed to space.

One possible problem with these choices for hyperboloids however
is that the area of this hyperboloid is infinite.  While for a
sphere a fixed radius area is $\int_{S^2} r^2 \sin\theta d\theta
d\phi= 4\pi r^2$, we have for a hyperboloid $\int_{H^2} r^2 \cosh
\theta d\theta d\phi =\infty$ for $r\neq0$. One point to note
then before doing the calculation is that the induced metric on
the connected hyperboloid is positive while the induced metric on
the two disconnect hyperboloids is negative. For the connected
hyperboloid, $H_2$, we have $x=a \cosh \theta \cos \phi$, $y=a
\cosh\theta \sin\phi$, $t=a \sinh\theta$.  For the disconnected
hyperboloids $H_1, H_3$ we have $x=a \sinh\theta \cos\phi$,
$y=a\sinh\theta \sin \phi$, $t=\pm a \cosh \theta$ and we also
notice that to cover the hyperboloids once we only need $\theta
\in [0,\infty)$. Calculating the area integral we have for one
disconnected hyperboloid and half of the connected hyperboloid
$-\int_0^\infty a^2 \sinh\theta d\theta d\phi + \int_0^\infty a^2
\cosh\theta d\theta d\phi=2 \pi a^2$. Note the minus sign
intuitively comes from the fact that one are is timelike and one
spacelike.  If we include the contribution from the other half we
get that the total area of the three hyperboloids is $4\pi a^2$!
This is exactly then an analytic continuation of a sphere's area.

Now let us consider a source, a delta function.  For the case of
Euclidean 3 dimensional space we have that integrating the flux
generated by a
delta function over a 2-sphere is $\int_{S^2} \frac{\hat{r}}{r^2}
\cdot \hat{r} r^2 \sin\theta d\theta d\phi= 4\pi$.  For the Wick
rotated case we have upon integrating the Green's function over
the three hyperboloids
\begin{eqnarray}
\left(\int_{H^2_1} + \int_{H^2_2} + \int_{H^2_3}\right)
\frac{\hat{\eta}}{\eta^2} \cdot \hat{\eta} \eta^2 (\cosh\theta
-\sinh \theta) d\theta d\phi=4\pi \ ,
\end{eqnarray}
where $\eta^2\equiv -t^2+x^2+y^2$. This verifies the consistency
of the definition of the delta function for spacelike sources with
the new definition of S-charge.

\subsection{Charge of the nonsingular time-dependent solutions}

Let us next try to apply this new rule of charge counting for our
non-singular imaginary sources.  The explicit calculation is
simplified by calculating at large values from the light cone.
Remember for two static charges at $x=\pm s$ that the potential is
\begin{equation}
\phi=\frac{1}{\sqrt{(x-s)^2+y^2+z^2}} +
\frac{1}{\sqrt{(x+s)^2+y^2+z^2}} \ .
\end{equation}
Although the configuration is non-spherical, the charge of this
system is easy to calculate for large values of $r^2=x^2+y^2+z^2$
where the solution simplifies to
\begin{equation}
\phi= \frac{2}{\sqrt{x^2+y^2+z^2}} + {\cal{O}}
\left(\frac{1}{r^2}\right)
\end{equation}
using a simple Taylor expansion.  To calculate the charge at
infinite radius, we see that the higher order terms do not
contribute and so the charge simply is $Q=2$.

We now try to similarly calculate the charge in the case time
dependent solutions with imaginary sources.  For example the Wick
rotation of the above case gives
\begin{eqnarray}
\phi &=&\frac{1}{\sqrt{(it-s)^2+y^2+z^2}} +
\frac{1}{\sqrt{(it+s)^2+y^2+z^2}} \nonumber \\
&=& \frac{\sqrt{2}}{((\eta^2+s^2)^2+4t^2s^2)^{1/4}}
\sqrt{1+\frac{\eta^2+s^2}
{\sqrt{(\eta^2+s^2)^2+4t^2s^2}}}.
\label{12imag+}
\end{eqnarray}
where $\eta^2\equiv |-t^2+y^2+z^2|$. For large $\eta \gg s$,
we have an expansion
\begin{eqnarray}
\phi&=&
\frac{2}{\eta} + {\cal{O}} \left(\frac{1}{\eta^2}\right) \ .
\end{eqnarray}
We see that this has a S-charge $Q=2$. Despite the fact that
there are neither real sources nor singularities, this
non-singular solution has a charge!  In this case the charge is
the same regardless of how close we are to the lightcone since
there are no real sources. On the other hand if the imaginary
sources have opposite charges, their S-charge is zero. This is
clear from Fig.~\ref{flip-wave-far} where we see a wave coming in
and flipping over. The contributions from the past and future
light cone regions clearly cancel to give zero S-charge.

\subsection{Lorentzian Gauss law}

It should be emphasized that in the case of the nonsingular
solutions, our notion of charge seems to be conserved only in an
asymptotic sense. Let us explain more concretely the deformations of
the integration surfaces.

In the usual definition of charges in Euclidean space, we are allowed to
deform the integration surface arbitrarily to enclose sources. This is
due to the Gauss's theorem.
A simple statement about Gauss's theorem in $1+2$ Lorentzian
spacetime is as follows.  Surround the origin with a cube with
sides at $\frac{1}{2}(\pm \Delta x_0, \pm \Delta x_1, \pm \Delta
x_2)$. For a scalar field $\phi$, the flux through the spatial
direction $x_2$ is
\begin{equation}
\nabla \phi\left(\frac{\Delta x_2}{2}\right) \cdot \hat{x}_2 \
\Delta x_1 \Delta x_0 + \nabla \phi\left(-\frac{\Delta
x_2}{2}\right) \cdot \left(-\hat{x}_2\right) \ \Delta x_1 \Delta
x_0
\end{equation}
which under Taylor expansion is
\begin{equation}
\partial_{x_2}^2 \phi(0) \ \Delta x_0 \Delta x_1 \Delta x_2 \ .
\end{equation}
The only difference for the flux calculation along the timelike
direction is that the unit vector has negative norm $\hat{x}_0
\cdot \hat{x}_0 =-1$ so the flux in the $x_0$ timelike direction
is
\begin{equation}
-\partial_{x_0}^2 \phi(0)  \ \Delta x_0 \Delta x_1 \Delta x_2 \ .
\end{equation}
In total the flux through the square is
\begin{equation}
(-\partial_{x_0}^2 +\partial_{x_1}^2+\partial_{x_2}^2)\phi(0) \
\Delta x_0 \Delta x_1 \Delta x_2
\end{equation}
so the generalized Gauss law in Lorentzian spacetimes
($\mu=0,1,2$) is simply
\begin{equation}
\int \partial_\mu \partial^\mu \phi \ d^3 x = \int \nabla \phi
\cdot \hat{n} \ d^2S \ .
\end{equation}

Instead of just restricting ourselves to using box-like
integration regions, let us further expand on the notion of
surrounding spacelike sources. A sphere surrounds a point
particle in Euclidean space, and a sphere in spacetime also
surrounds a spacelike source. Arbitrary deformation of the
bounding surface are allowed as long as we do not pass through
additional charges.  One can imagine then a large deformation of
the bounding surface where the edge stretches far from the
source.  If the limit in which the contour deformation is taken to
infinity carefully, then the charge bounded must remain constant.

For example in the case of the singular solution, there is a
concrete source producing an associated charge. For the
non-singular sources there is always an associated charge as
calculated using our prescription. However if deformations of the
hyperboloids to a sphere are smooth, then we take the resulting
sphere to zero radius and the charge accordingly vanishes due to
the smoothness of our solutions. Therefore we should understand
that our prescription of using hyperbolas for the S-charge is not
necessarily topologically equivalent to the sphere which can
shrink. The integration surfaces consist of three hyperboloids in
the case of 1+2 dimensions, and they are disconnected, that is
why we obtained nonzero S-charge for non-singular field
configurations.

\subsection{Comparison with previous charge definitions}

Let us discuss relations to the previous definitions of spacelike
charges. A definition of spacelike charge for the case of 1+1
dimensional Klein-Gordon was presented as an example in
Ref.~\cite{stro}.
As discussed there
one choice of propagator is
$\phi(u=t+x,v=t-x)=\theta(u) \theta(v)$ as it satisfies
\begin{equation}
\partial_u \partial_v \phi(u,v)=\delta^2(u,v) \ .
\end{equation}
This solution satisfies a straightforward but
probably unfamiliar Gauss law equation which is still of the form
$\int_V d*d\phi=\int_{\partial V} *d\phi$ but where the integral
is now over the space and time direction. First let us calculate
the S-charge of this configuration using the definitions of
Ref.~\cite{stro} where we enclose the spacetime origin. If we
integrate over a rectangle in lightcone coordinates with corners
at $(u,v)=(\pm a,\pm b)$, we calculate the charge of this
solution to be
\begin{eqnarray}
Q=\int\!\! * d\phi=
\oint\! \left[\delta(u) \theta(v) du -\delta(v) \theta(u) dv\right]
=\int_{-a}^a\!\!\!\!\! \theta(b) \delta(u) du -
\int_b^{-b}\!\!\!\!\!\!\!\theta(a) \delta(v) dv=2 \ .
\nonumber
\end{eqnarray}
If we take this configuration to be a solution to Maxwell theory
in 1+2 dimensions, we must also impose the current conservation law
$\partial_t \rho +
\partial_x J_x=0$ which introduces the spacelike current
$J_x=-\theta(x)\delta^\prime(t)$.

Let us compare this result with our prescription for the S-charge.
In this 1+1 dimensional case, four
disconnected hyperbolas are needed to bound the source.  The
propagator $\Phi=\theta(u)\theta(v)$ can be written in terms of
Milne $(t,x)=\pm(T\cosh r, T\sinh r)$ and Rindler coordinates
$(t,x)=\pm (-T \sinh r, T\cosh r)$.  In the Milne region $t>|x|$ we
calculate the charge to be
\begin{eqnarray}
\int_M * d\Phi= \int [-\delta(Te^r) \theta(Te^{-r}) e^r T
-\delta(Te^{-r}) \theta(Te^r) e^{-r} T]dr
\nonumber \\
\hspace{20mm}+ [-\delta(Te^r) \theta(Te^{-r}) e^r+\delta(Te^{-r})
\theta(Te^r) e^{-r}] dT \ .
\end{eqnarray}
For simplicity take the integral to be at fixed $T$ and from
$r=(-\infty, \infty)$ and change variables to obtain
\begin{equation}
\int_M * d\Phi=\int_0^\infty -\delta(w) \theta(T/w) dw - \delta(w)
\theta(T/w) dw =\frac{1}{2}\theta(T \cdot \infty)
+\frac{1}{2}\theta(T\cdot \infty)=1
\end{equation}
where we have used the identity
\begin{eqnarray}
\int_0^\infty \delta(x) dx=1/2.
\label{delta}
\end{eqnarray}
Although the delta function is sharply defined at $x=0$, we define
by the meaning of our hyperbolic integral this symmetric condition
on the integration of the delta function.   In the second Milne
region $t<-|x|$ we find that the answer is zero $Q=0$. This can be
seen simply since we can obtain this second region from the first by
simply replacing $T\rightarrow -T$. The final charge will then
depend on $\theta(-T\infty)=0$ giving zero charge.

For the Rindler region $x>|t|$ we find the charge integral becomes
\begin{eqnarray}
\int_R * d\Phi=\int [\delta(Te^{-r}) \theta(-Te^r) e^{-r} T
-\delta(-Te^r) \theta(Te^{-r}) e^{r} T]dr
\nonumber \\
\hspace{20mm}- [\delta(Te^{-r}) \theta(-Te^r) e^{-r}+\delta(-Te^r)
\theta(Te^{-r}) e^{r}] dT \ .
\end{eqnarray}
Again integrating at fixed $T$ from $r=(\infty,-\infty)$ and using
our above definition of integrating delta functions
\begin{equation}
\int_R * d\Phi=\int_0^\infty \delta(w) \theta(-T/w) dw
-\delta(w)\theta(T/w) dw=\frac{1}{2}\theta(T\cdot
\infty)=\frac{1}{2}
\end{equation}
we get a charge contribution of $1/2$. Similarly in the second
Rindler region the answer is $1/2$. In total, the charge of this
spacelike source is $Q=2$ just like when we integrated before. In
this case we have defined the hyperbolas to bound the charge
which results in the same charge.

Note that the charge using the hyperbolas would take a different value
if we used a different way of evaluating the boundary of the integrals.
Here we took only half of the delta function as in (\ref{delta}),
but if we define the integral as
$\lim_{\epsilon \to 0} \int_\epsilon^\infty dw$ then the resulting
S-charge is zero.
With this definition of the boundary with $\epsilon$,
the hyperbolas of the integral never touch the
lightcones and so with this definition it does not enclose the origin
where the source resides.

Before ending this subsection, we evaluate the charge of the Wick
rotated logarithm harmonic function in two dimensions using the
definition of Ref.~\cite{stro}. The solution to the Green's
function in 2 dimensional Euclidean space is known to be
$(1/4\pi)\ln(x^2+y^2)$ and its Wick rotation $(1/4\pi)\ln
(x^2-t^2)$ was the basis of our discussion in the previous
section. Although we have referred to this kind of solution as a
delta function for the wave equation due to analytic
continuation, this kind of procedure is a bit subtle in the
presence of delta functions. Let us therefore verify the form of
singularity by applying Gauss's theorem as before.  To ensure
that Wick rotation is consistently working we check the charge of
this Wick rotation using once again light cone coordinates
$(u,v)$.  Taking into account the $i\epsilon$ prescription which
rewrites integrals in terms of the principal value and delta
functions we have
\begin{eqnarray}
&&*d \ \ln(x^2-t^2)= \frac{v du- u dv}{uv+i\epsilon}\\
&&Q= \int
\left[
vdu \left(P\left(\frac{1}{uv}\right) -i\pi \delta(uv)\right) -udv
\left(P\left(\frac{1}{uv}\right) -i\pi \delta(uv)\right)
\right]
\nonumber\\
&&\hspace{5mm}= \int
\left[
-i\pi \ {\rm Sign}(v) \delta(u) du + i\pi \
{\rm Sign}(u) \delta(v)
dv\right]=4\pi i
\end{eqnarray}
so the solution $\phi=(i/4\pi) ln(x^2-t^2)$ has a unit of
S-charge.

\subsection{Relation to usual static charge}

\begin{figure}[t]
\begin{center}
\includegraphics[width=12cm]{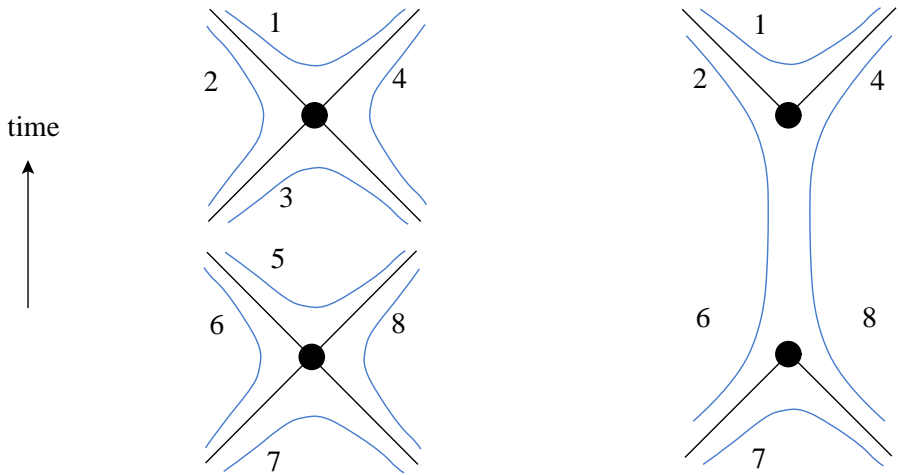}
\caption{Surround two spacelike source using hyperbolas, and then
cancel contributions whenever possible.  For example contributions
from hyperbolas 3 and 5 cancel.
After making all possible cancellations we are left with a
similar configuration as for one S-brane.} \label{2Scharges}
\end{center}
\end{figure}

In this subsection we count more complicated configurations of
charge. First take the case of two S-branes shifted in time as
discussed further in Sec.~\ref{timearray}. Our rule is to surround
both S-branes by a single large hyperboloid as shown in
Fig.~\ref{2Scharges}.
These hyperbolas have cancelling contributions and we can deform
the calculation into three hyperboloids again as shown in
Figure~\ref{2Scharges}.  The calculation for two S-charges is
similar to that of one S-charge.  In the process we must
integrate over the lightcone discontinuities.

\FIGURE[t]{ \epsfig{file=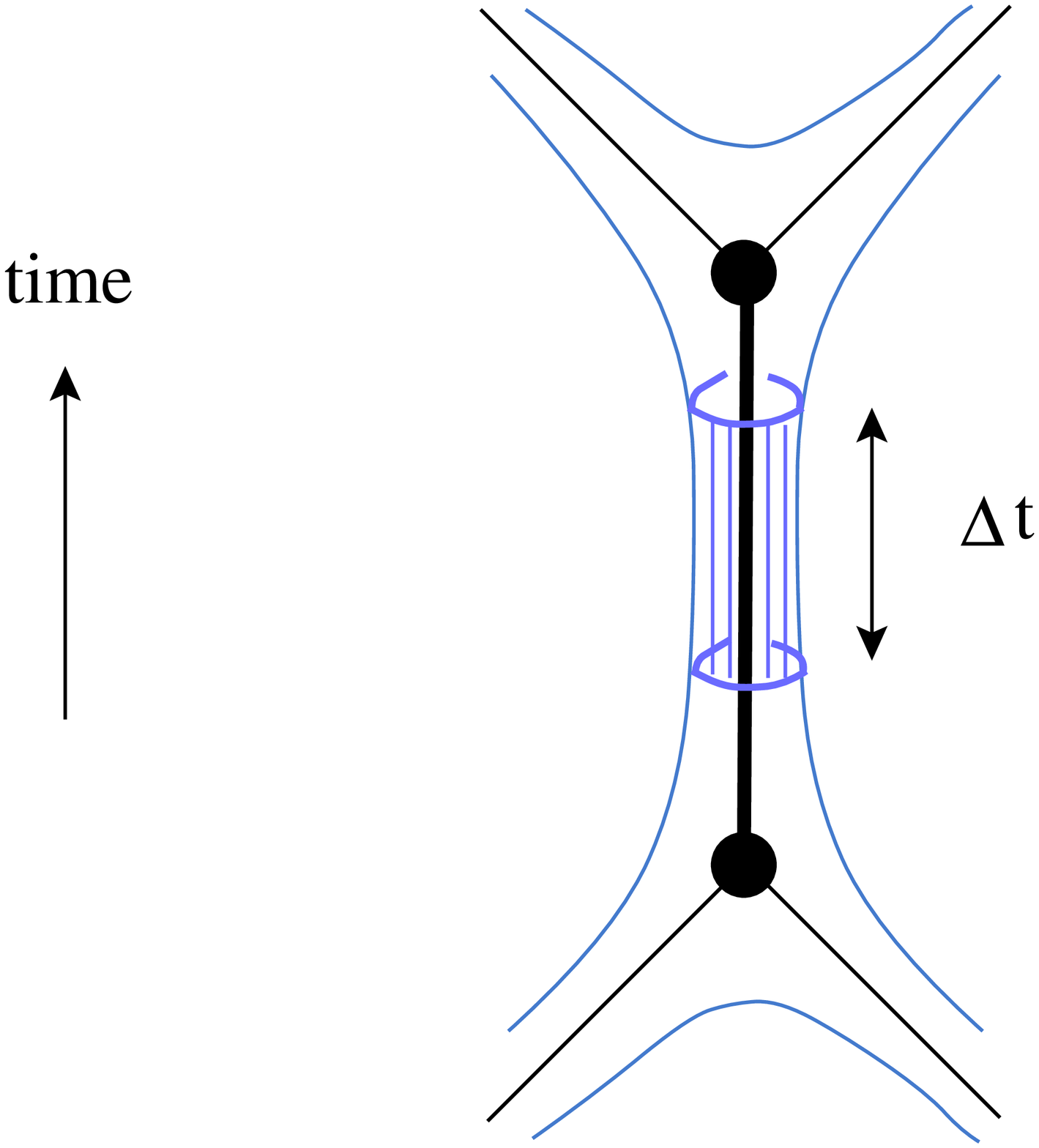,width=5.5cm}
\caption{The integral over a small region of a hyperbola turns
into an integral over a cylinder with the circle in the space
directions and the length of the cylinder along time.}
\label{Scharge-charge} }
A further interesting observation which shows the consistency of
our definitions is the following.  Allow for a linear
distribution of singular spacelike sources in the time direction
as we will discuss in Sec.~\ref{timearray}. Generalizing the
definition of S-charge just outlined above for two charges then
we should integrate over three disconnected regions. If the
S-brane is long lived then two of the regions are past the
singularities and do not contribute much to the total S-charge.
We are just left with then a circular integral of the usual
charge for a particle, times an integral over time as shown in
Fig.~\ref{Scharge-charge}
\begin{equation}
Q_{\textup{S-charge}}= Q_{\textup{static charge}} \cdot \Delta t\
.
\end{equation}

Our definition of S-charge is then related to the usual charge of
a static particle charge in the same way as static charge is
related to a linear charge density in space, $Q=\lambda L$.  In
this way our definition of S-charge reproduces the usual static
charge.  Our S-charge is therefore also consistent with the fact
that integrating singular spacelike sources over time reproduces
the potential of an static charged particle as we will mention in
Sec.~\ref{timearray}. Because spacelike sources can be thought of
concretely as the individual bits of a particle's worldline
trajectory, there is something quite fundamental about spacelike
sources in time dependent systems as opposed to particles.

\subsection{Secondary S-charge}

In addition to the above definition of S-charge, there has also
been a definition of S-brane charge proposed in Ref.~\cite{greg1}.
First let us review this definition.  Turning away from the
Klein-Gordon equations, consider the a positive charge and a
negative charge forming a dipole. There are three charges we can
calculate which are the dipole charge and also each individual
charge.  Place a negative/positive charge at $z=\pm s$ so the
potential is
$\phi=\frac{c}{\sqrt{x^2+y^2+(z-s)^2}}-\frac{c}{\sqrt{x^2+y^2+(z+s)^2}}$.
We surround the negative charge by a two sphere which we deform
until it fills out the half plane between the two charges and goes
out to infinity in the upper half plane.  The electric field along
this $z=0$ plane is
\begin{eqnarray}
E_z=\frac{-c(z-s)}{(x^2+y^2+(z-s)^2)^{3/2}}
+\frac{c(z+s)}{(x^2+y^2+(z+s)^2)^{3/2}}
=\frac{2cs}{(x^2+y^2+s^2)^{3/2}}.
\end{eqnarray}
If we change variables so $\rho^2=x^2+y^2$ then the flux through
the half plane is
\begin{equation}
\Phi=\int E_z \rho d\rho d\phi=2\pi \int
\frac{2cs}{(\rho^2+s^2)^{3/2}} \rho d\rho=2\pi(2c s)
\frac{1}{\sqrt{\rho^2+s^2}}\biggm|^\infty_0 =-4\pi c\end{equation}
which is the same as the calculation using the usual sphere. Such
a charge definition works only on two opposite charges and is
zero when the charges are the same. This method picks up one of
the charges but only in a dipole configuration. The flux
calculation is of course the same for any choice of infinite half
plane between the two charges; the only difference is that there
will be also be contributions from the electric field in the
other $(x,y)$ spatial directions.

Wick rotating the two oppositely charged sources by $z\rightarrow i
\tau$ does not affect the above charge calculation.  What we have
though is a different interpretation of the charge. The potential
$\phi=A_0\rightarrow A_4$ and the original electric field along the
$z$ direction becomes an electric field along the $x^4=i \tau$
direction so overall $F_{tz}=F_{x^4 \tau}$. Specifically the
electric field along $x^4$ at time $\tau=0$ is
\begin{equation}
E_{x^4}=\frac{2cs}{(x^2+y^2+s^2)^{3/2}}
\end{equation}
and the rest of the calculation is the same.  Calculating the
flux at a different time gives the same answer but the difference
is that now there is also the presence of time dependent magnetic
fields.  If we were to calculate the charge of such a potential
in a scalar field theory, we would obtain zero S-charge according
to our prescription since this is a wave which flips its
amplitude.

Summarizing the difference between the definitions of S-charges,
we find that our new definition of S-charge seems consistent with
the notion of calculating the flux from putting in a single delta
function sources in time. It is also able to calculate the total
amount of charge which is focused into a small region of space
and time.  The notion of S-brane charge in Ref.~\cite{greg1} on
the other hand is a way to calculate the charge of dipole type
configurations. These two charge definitions produce different
results in the cases we have studied. However it is very natural
for spacelike sources to have various charges.  We are well
familiar with the fact that distributions of electric charges
have monopole, dipole, quadropole, and higher multipole moments
and that is something similar to the situation here.

%%%%%%%%%%%%%%%%%%%%%%%%%%%%%%%%%%%%%%%%%%%%%%%%%%%%%%%%%%%%%%%%%%%
%%%%%%%%%%%%%%%%%%%%%%%%%%%%%%%%%%%%%%%%%%%%%%%%%%%%%%%%%%%%%%%%%%%
%%%%%%%%%%%%%%%%%%%%%%%%%%%%%%%%%%%%%%%%%%%%%%%%%%%%%%%%%%%%%%%%%%%
%%%%%%%%%%%%%%%%%%%%%%%%%%%%%%%%%%%%%%%%%%%%%%%%%%%%%%%%%%%%%%%%%%%
%%%%%%%%%%%%%%%%%%%%%%%%%%%%%%%%%%%%%%%%%%%%%%%%%%%%%%%%%%%%%%%%%%%%
\section{Explicit Examples}
\label{sec:more}

In this section we provide explicit examples of the imaginary
source solutions for massless/massive Klein-Gordon systems and
also for Maxwell system. In particular, Wick rotations of
electrons, monopoles and instantons are shown to give
cylindrical/spherical electromagnetic waves.

\subsection{Arrays of massless Klein-Gordon solutions}

\subsubsection{Infinite imaginary arrays} \label{infinitearraysec}

\noindent\underline{Arrays in 4 dimensions}
\vspace{2mm}

In Euclidean space, let us take an infinite array of alternating
point charges.  The potential is
\begin{equation}
\phi_{\rm array}=\sum_{n=-\infty}^{\infty}
\frac{C_1}{(x_4-s/2+2ns)^2 + r^2} + \frac{C_2}{(x_4+s/2+2ns)^2
+r^2}
\end{equation}
where $r^2\equiv (x_1)^2+(x_2)^2+(x_3)^2$. We have been dealing
with finite numbered arrays but what about infinite numbered
arrays like the one above? Does it even give a well defined
potential when we are away from the point charges?  Although an
infinite array may seem like a very special configuration, it is
important to discuss this case especially in relationship to the
rolling tachyon.

Let us check that $\phi_{array}$ is bounded when away from a
single charge, for the case of real $C_1=C_2$.
First it clear that the denominators are squares
of positive numbers and so will be minimized when each positive
number is as small as possible.  This means that we should take
$r=0$.  Due to the periodicity of the solution we know that we
can maximize the solution by simply maximizing it in the range
$-s/2<x_4<s/2$.  In this range all terms with $n\neq 0$ are
maximized at $x_4=\pm s/2$.
Summing all such non-singular terms the sum
is bounded by
\begin{equation}
\phi_{\rm array}<\sum_{n=-\infty}^{\infty} \frac{2}{(2ns)^2} +
\frac{1}{(x_4-s/2)^2} + \frac{1}{(x_4+s/2)^2} \ .
\end{equation}
The infinite sum is bounded and the remaining two terms are only
divergent at the singularity.  By taking absolute values, we can
likewise show that arrays of positive and negative charges
($C_1 = -C_2$) give a
bounded sum for the potential except for when we are at a source.

In the case of time dependent solutions, we can similarly bound the
potential.  One easy way to check this statement is to realize that
the potential for two positive charges was seen to maximized at the
origin of spacetime (see Fig.~\ref{fig1}).  This means that for each
pair of same charges in the infinite sum the same is true that the
potential is maximized at $x_0=0, x_1=x_2=x_3=0$. This again reduces
the sum to an infinite series of inverse squares which is finite.
The difference is that in this case, all the terms are non-singular
since we can never approach any of the singularities at $x_0=\{\rm
imaginary \ values\}$. We conclude that the potential is completely
bounded.  As pointed out in Ref.~\cite{sthermo}, this infinite summation
can be done explicitly for $C_1=C_2=C$, using the relationship\
\begin{equation}
\tanh z=\lim_{n \to \infty} \sum_{k=-n}^n \frac{1}{i\pi
(\frac{1}{2}-k) +z} \qquad\mbox{for }\quad \frac{iz}{\pi}-\frac{1}{2}
\notin {\bf Z}
\end{equation}
we find the non-singular form
\begin{eqnarray}
\phi_{\rm RT}&=&\sum_{n=-\infty}^\infty
\frac{C}{-(x_0-is/2+2ins)^2 + r^2}
+ \frac{C}{-(x_0+is/2+2ins)^2 +r^2}\nonumber\\
&=&\frac{1}{r}\frac{C\pi}{2s} \left[\tanh \left(\frac{\pi}{s}
(-x_0+r)\right) + \tanh \left(\frac{\pi}{s} (x_0+r)\right)\right]
\label{rollingtachyon}
\end{eqnarray}
At time $x_0=0$ this function is monotonically decreasing as a
function of the radius $r$ from the value 1 to zero.  This can be
interpreted as a localized lump.  For large values of the radius
$r\gg(x_0,s)$, we have that this is a nearly static potential.
Furthermore, this has a characteristic behavior $\phi_{\rm RT} \sim
1/r$ as opposed to the original $1/r^2$ contribution from each
source.  In the far past and far future, this potential goes to
zero.
\FIGURE[t]{
\epsfig{file=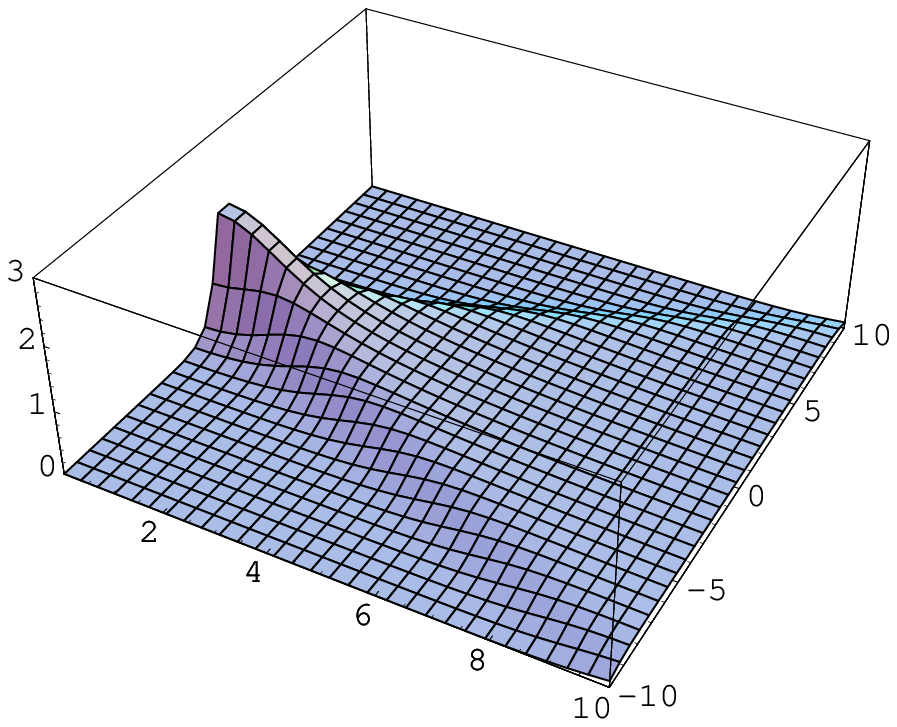,width=7cm}
\caption{An infinite array of like codimension four charges is
asymptotically in the spacelike direction a codimension three
particle and a decaying wave in the past and future.  The lightcone
clearly separates the characteristics in the two different regions.}
\label{Strominger-array-far}
}

Here we see that arrays of imaginary sources of codimension four,
can create time dependent versions of real sources of codimension
three. In effect we are creating a nonsingular time dependent
version of a static point source!  This can be understood from
the fact that an infinite summation of sources is effectively an
integral and so the $1/r^2$ behavior sums to $1/r$. Imaginary
source summation is not effectively a summation over time but a
summation over space because of the ``$i$'' imaginary factor.
Therefore essentially any infinitely extended in imaginary space
summation of one type of charge will lead to the same kind of
asymptotic $1/r$ behavior at spatial infinity.\footnote{We note
in that gravity, the imaginary source solution corresponding to
diholes \cite{greg1,cards1}, as opposed to an infinite number of
solutions in this KG field theory case, was sufficient to
generate a conical defect.}

A peculiar fact about such infinitely extended solutions is that
at fixed time and large radius, these solutions have finite static
charge. By ``static charge'' here we mean integrating the three
form $*d\phi$ over the time direction and angular directions at a
fixed radius, $\int_{\Delta t, \Omega} *d\phi$.  Next divide the
integral by the period of time we integrate over and this
averaged result is what we mean by the static charge.
If we take the limit of the radius to infinity first, then the
static charge is a finite constant.  Therefore an observer would
expect to see an electron at the origin.  However over a long
enough period the time dependent terms have a dramatic effect and
are not ignorable in general.  The effect of time dependence is
to transport the field configuration to and from infinity.

While we discussed infinite arrays of one type of charge, we can
also discuss what happens to alternating charge arrays.  In fact
only these dihole charge type solutions have known gravity
solution duals \cite{greg1, cards1}.  It is actually a simple
process to produce the infinite array of such solutions, since
the sum of alternating charges can equivalently be thought of as
the sum of two infinite arrays of like charges.  Simply take the
above solution and translate $x_0\to x_0+is/4$ and take the
imaginary part of the result. Using the standard hyperbolic
tangent addition rule we get\cite{sthermo}
\begin{equation}
\phi_{\rm RT2}=\frac{1}{r}\frac{C\pi}{s} \left[
\frac{1}{\cosh(2\pi(r+x_0)/s)} -
\frac{1}{\cosh(2\pi(r-x_0)/s)}
\right]
\end{equation}
which is clearly going to zero at large radial values and in the
far past and future.

One can also modify the charges $C_1$ and $C_2$ so that they are
dependent on $n$. As long as the modification is mild (the charge
does not grow as $n$), the final sum is convergent and well-defined.
This variation corresponds to various fine structure of the
spherical shell shrinking and expanding.

\vspace{10mm}
%%%%%%%%%%%%%%%%%%%%%%%%%%%%%%%%%%%%%%%%%%%%%%%%%%%%%%%
\noindent\underline{Arrays in 1 dimension}

Each source localized in one dimension generates a linear potential,
$\phi=|x-a|$, which has a derivative discontinuity.  Although taking
an array of such sources is a bit subtle due to the infinities
involved, a consistent representation is the triangle wave function.
Placing negatively charged sources at $x=s/2+2ns$ and positively
charged sources at $x=-s/2+2ns$ for all integers $n$, then the
potential is $\phi=2x/s$ for $x=[0,s/2]$, $\phi=2-2x/s$ for
$x=[s/2,3s/2]$ and $\phi=-4+2x/s$ for $x=[3s/2,2s]$, and so on.
This function
periodically repeats with period $2s$. The derivative
discontinuities occur at the sources of the charge and alternate
sign.  This has a well known Fourier representation
\begin{equation}
\phi(z)=\frac{8}{\pi^2} \sum_{n\ \textup{odd}}
\frac{(-1)^{(n-1)/2}}{n^2} \sin\left(\frac{n\pi x}{s}\right) \ .
\end{equation}
We can then easily shift $x$ and Wick rotate to get what appears to
be an infinite sum of Sen's rolling tachyons
\begin{equation}
\phi(z)=\frac{8}{\pi^2} \sum_{n\ \textup{odd}}
\frac{(-1)^{(n-1)/2}}{n^2} \cosh\left(\frac{n\pi t}{s}\right) \ .
\end{equation}
Plotting this solution as a function of $t$ and for finite $n$ it
looks approximately linear near the origin before exhibiting
exponential growth.  The exponential growth will begin at earlier
times as we add more and more terms so in the limit of an infinite
number of terms there is no linear region.  In fact we can easily
show that this infinite series has a zero radius of convergence
due to the hyperbolic terms growing too rapidly as compared to the
$1/n^2$ suppression factors in the coefficients.\footnote{
Shifting and analytically continuing $x$ results in an infinite
sum of hyperbolic sines which does not help the convergence
properties. Also taking the sources to all have the same charge
does not help and in fact the potential does not appear well
defined in this case.}

One might note a particular feature of this analysis in that if we
extrapolate the Wick rotation of the Fourier expansion for all
periodic arrays of solutions, then the Wick rotated solution is
apparently badly divergent unless there is a very special
cancellation of the exponentially growing modes. So while
\begin{equation}
f(x)=\sum_n a_n e^{2\pi inx}
\end{equation}
is a reasonable expansion for periodic static functions, the Wick
rotation of this function is not necessarily well represented by the
Wick rotation of the Fourier expansion in the case of infinite sums
being needed. If we Wick rotate $x\rightarrow i t$, then we have the
expansion in terms of exponentially growing modes
\begin{equation}
f(it)=\sum_n a_n e^{-2\pi nt}
\end{equation}
which is certainly not absolutely convergent and in fact leads to
divergences. However for specially chosen coefficients $a_n$, this
sum does converge for a certain radius: an example is $a_n =
e^{-n^2}$.

As we already know Wick rotations of infinite arrays such as
the one in 4 dimensions we saw are finite.  In the present case
analytic continuation of the Fourier sum is not reliable and does
not provide us a proper representation of the Wick rotated function.
The Wick rotations of a solution and a representation of that
solution in some basis are not necessarily consistent.
In showing the completeness in $1+0$ dimension, we have been careful on
this and took $s\to\infty$ limit to avoid the confusion.

\subsubsection{Arrays in real time} \label{timearray}

In addition to arranging pairs of singularities at one imaginary
time, we can also array them in time.  This is very simple since for
example whenever we encounter terms of the form $-(t-is)^2 +
(x_1)^2+(x_2)^2+(x_3)^2$, we can also simply translate this in real
time by $t\rightarrow t+t_0$.  The case of two and three such
configurations are shown below in Fig.~\ref{2nonsing-Sbranes}.

\FIGURE[t]{
\epsfig{file=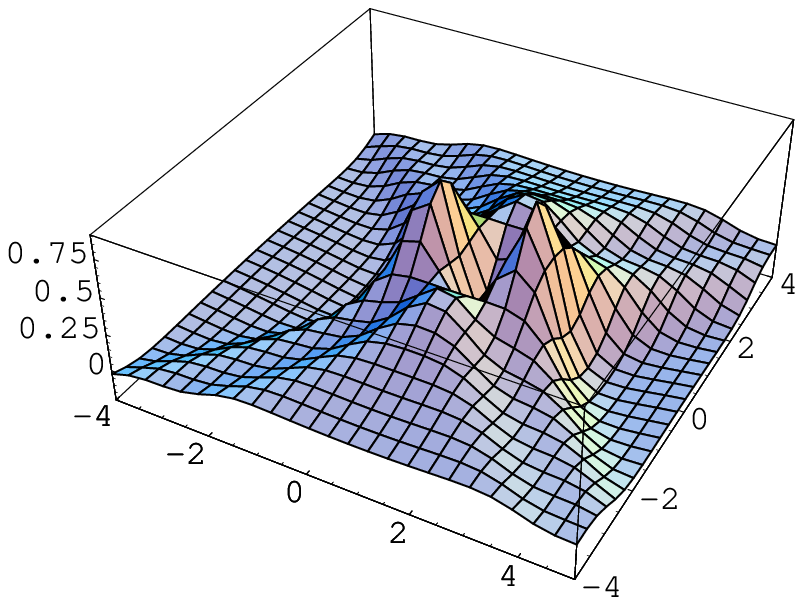,width=7cm}
\epsfig{file=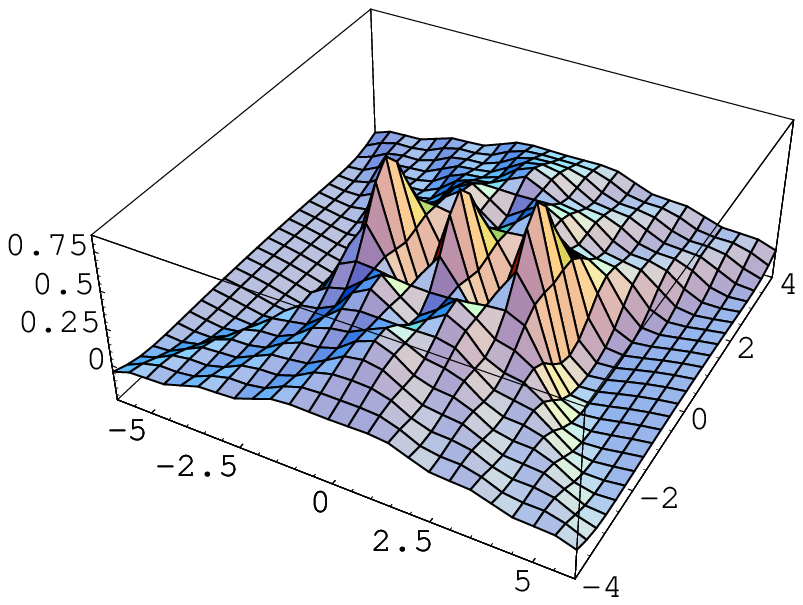,width=7cm}
\caption{
Two (left) / three (right)
non-singular pairs of S-branes parallel in time.}
\label{2nonsing-Sbranes}
}

Next we ask what happens if we arrange S-branes continuously in
time.  If we arrange the singular S-branes sources in time then we
can create the usual static objects in this way. This is simple to
see from the KG example.  If we have the equation
\begin{equation}
(-\partial_t^2 + \partial_1^2 + \p_2^2+ \p_3^2) \phi = \delta(t-t_0)
\delta(x_1)\delta(x_2)\delta(x_3)
\end{equation}
and integrate both sides with respect to all values of $t_0$,
then the right hand side becomes the three dimensional spatial
delta function.  This will also automatically produce the usual
harmonic solution $\Phi=\int\! dt \ \phi=1/4\pi \sqrt{\vec{x}^2}$
to the Laplacian equation in three dimensions. In other words
integrating an S-brane in time produces a pointlike static
solution. It is also possible to create sources that begin and
end in time, which are singular S-branes existing for finite
time. Like the S-branes which exist for a moment in time, finite
sources create light cone discontinuities starting and ending on
the endpoints of the S-branes.

Having discussed the singular case, what static solutions we
produce upon integrating imaginary sources in time? Take the four
dimensional case where we must integrate
\begin{equation}
\int_{-\infty}^{\infty} \frac{dt_0}{-(t-t_0-is)^2 + (x_1)^2+
(x_2)^2+(x_3)^2}. \label{zerointegral}
\end{equation}
The result is zero, for both of our four dimensional non-singular
imaginary source solutions (\ref{sol1phi}) and (\ref{sol2phi}).
This is technically because, for example for the latter solution,
it takes both positive and negative values and integrating it over
time gives zero.  In addition this integral must be zero since
otherwise we would have a non-singular static localized solitonic
solution in four dimensions which is known to be impossible. Our
result is therefore consistent. However as long as the solution
is not infinitely extended in time (that is, the integral limits
in (\ref{zerointegral}) for the time-dependent solution are
finite) then we generate an arbitrarily long-lived solution.

Another useful example is the partial integration of the ``rolling
tachyon'' solution (\ref{rollingtachyon}) which we take without
loss of generality from $-t_1$ to $t_1$
\begin{eqnarray}
\int_{-t_1}^{t_1} \!\!dt_0 \;\phi(r,t-t_0)
&=&\int_{-t_1}^{t_1} \!\!dt_0 \;\frac{1}{r}\frac{C\pi}{2s}
\left(\tanh \frac{\pi(r-t+t_0)}{s} + \tanh
\frac{\pi(r+t-t_0)}{s}\right)
\nonumber \\
&=&\frac{C}{2r} \log\left[
\frac{\cosh [2\pi (r+t_1)/s]+\cosh [2\pi t/s]}
{\cosh [2\pi (r-t_1)/s]+\cosh [2\pi t/s]}\right].
\end{eqnarray}

\FIGURE[t]{ \epsfig{file=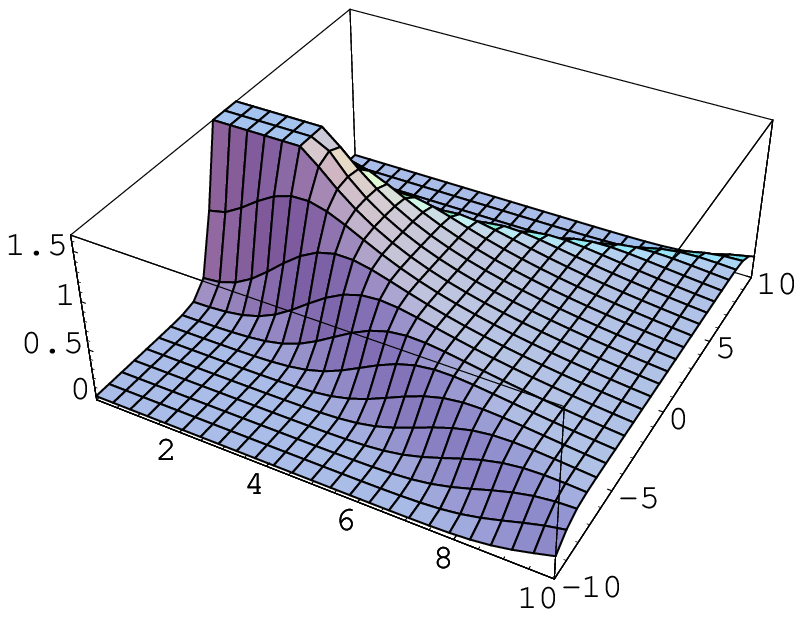,width=7cm}
\epsfig{file=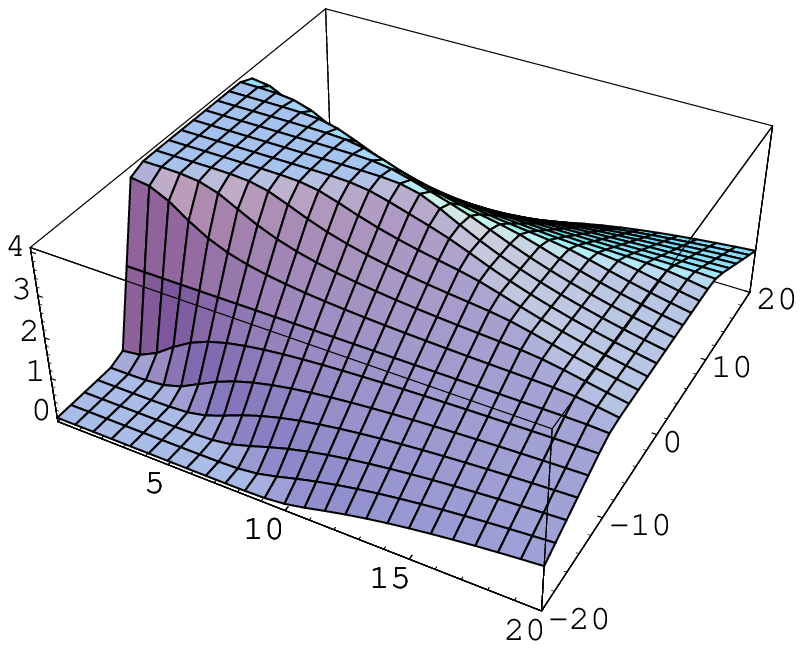,width=7cm} \caption{Left:
Integrating the infinite array for a short period of time,
$-5<t<5$, gives a solution similar to the un-integrated version.
Right: Integrating the infinity array for $-10<t<10$, we find that
the solution develops a nearly constant region near the origin of
spacetime. } \label{integrateRT}}

The spatial asymptotics are still like the un-integrated infinite
array.  As the array is integrated over longer periods of time,
the potential near the origin becomes a constant as shown in
Fig.~\ref{integrateRT}.

\subsection{Massive Klein-Gordon system}

For a more explicit discussion, we here provide imaginary source
solutions to massive Klein-Gordon theories in 1+1 and 1+2
dimensions.

For the Euclidean massive Klein-Gordon operator, we know the
Green's function in 3 dimensions
\begin{eqnarray}
 \phi = \frac{qe^{-mr}}{4\pi r}
\end{eqnarray}
is a solution of the equation of motion with a source,
\begin{equation}
(\partial_i^2 -m^2) \phi=q\delta^3(x_i), \quad (i=1,2,3),
\end{equation}
where $m$ is the mass and $r$ is a radial variable in the 3
dimensional space.  This can be used for getting a time-dependent
solution in 1+2 dimensions. We can go ahead for getting imaginary
source solutions by shifting the location of the source and
performing the Wick rotation,
\begin{eqnarray}
 \phi = q \frac{\exp[-m\sqrt{(-it+s)^2+ x^2+y^2}]}
{\sqrt{(-it+s)^2+ x^2+y^2}}.
\end{eqnarray}
For example, if we take the real part of the above solution,
or in other words, we consider like charged sources located at
$t=\pm s$, then we get a time-dependent solution
\begin{eqnarray}
 \phi = q\frac{e^{-mA}}{A^2+B^2}
\left[A \cos(mB)-B\sin(mB)\right]
\end{eqnarray}
where $A$ (or $B$) is the real (or the imaginary) part of
$\sqrt{(-it+s)^2+ x^2+y^2}$,
\begin{eqnarray}
&& A = \sqrt{(\Delta + \sqrt{\Delta^2 + 4s^2 t^2})/2}, \quad
 B = -st/A, \quad
 \Delta = -t^2 + x^2 + y^2 + s^2.
\hspace{5mm}
\end{eqnarray}
When an observer is spatially far from the light cone, i.e. when
$-t^2+x^2 + y^2 \gg s^2$, then we may approximate $A \sim
\sqrt{-t^2+x^2+y^2}$, so the solution looks like $\phi \sim q
\exp [-m\sqrt{-t^2+x^2+y^2}]/\sqrt{-t^2+x^2+y^2}$. On the other
hand, when $x^2+y^2 \ll t^2-s^2$, we have $A \sim s$ and so
$B\sim -t$. Thus the solution is oscillatory, $\phi \sim q
\exp[-ms]\sin[-mt](1/t)$.  Finally, if we take $s\gg (x,y,t)$ then
the solution simplifies to the stationary wave $cos(mt) \
(e^{ms}/s)$.   In sum, inside the light cone the solution is
oscillating in time and thus looks almost like a plane wave with
zero velocity, while outside of the light cone the solution is
exponentially decaying asymptotically with length-scale given by
proper distance from the lightcone, $\sqrt{-t^2+x^2+y^2}$.  From
the large $s$ limit we can see straightforwardly that imaginary
sources will generate all the sourceless 1+2 massive Klein-Gordon
solutions as all such solutions arise by Lorentz transformations
of the stationary wave.

In the case of 1+1 dimensions, we start again by examining the Green's
function in 2 Euclidean dimensions. It
has the modified second Bessel function solution
\begin{equation}
\phi(x,y)=-\frac{1}{2\pi} K_1(-m\sqrt{x^2+y^2})
\end{equation}
which is exponentially decreasing at large radius and goes to
infinity at the origin.  Next take two such sources at $y=\pm is$
and perform the same Wick rotation $y \to i t$.  Taking the real and
imaginary parts of the solution to get
\begin{eqnarray}
\phi_{\rm R}
& = & -\frac{1}{2\pi} {\rm Re}\left[K_1(-m\sqrt{x^2-(t-is)^2})\right], \\
\phi_{\rm I} & = &
-\frac{1}{2\pi} {\rm Im}\left[K_1(-m\sqrt{x^2-(t-is)^2})\right],
\end{eqnarray}
we find nonsingular time dependent solutions to the massive Klein
Gordon equations of motion as shown in Fig.~\ref{mKG-pictures}.
\FIGURE[t]{ \epsfig{file=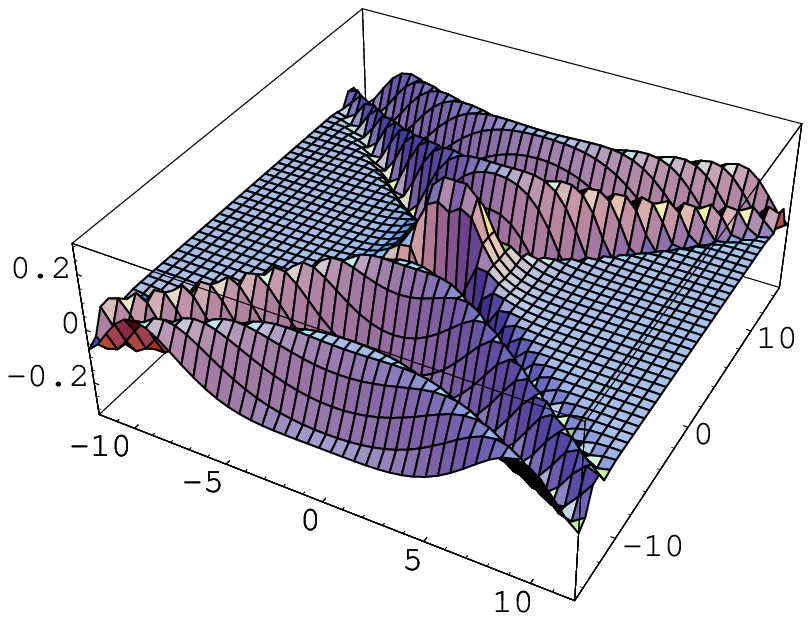,width=7cm}
\epsfig{file=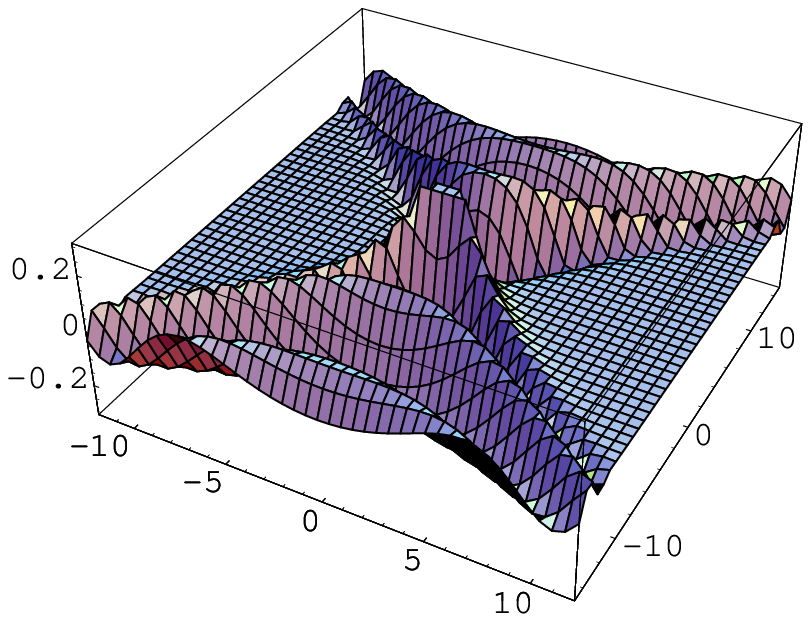,width=7cm}
\caption{Left: 1+1 KG solution for two imaginary source at $t=\pm
i$ and with the mass set to one, $m=1$. The ripples are in the
forward and past light cones and create a lump at $t=0$. Right:
Solution for two oppositely charged imaginary source at $t=\pm
i$. The ripples are in the forward and past light cones and
create a flat zero solution at $t=0$. } \label{mKG-pictures} }

These solutions are asymptotically flat at large spatial
directions and have damped oscillatory behavior in the past and
future light cone.  This oscillatory behavior is qualitatively
different from the massless solutions with imaginary sources and
reminiscent of the minisuperspace approximation for S-branes in
string theory discussed in Ref.~\cite{OScreation}. As with the
massless KG case, these solutions are symmetric in space and
either symmetric or antisymmetric in the time direction.

In the case of 1+0 dimension, a generic solution has been presented in
Sec.~\ref{sec1+0}.

\subsection{Maxwell system}
\label{secmaxwell}

In this subsection we consider the Maxwell equations, another
important physical system. We restrict our attention to 1+3 or 4
dimensional space(time).
The Euclidean equations of motion are
\begin{eqnarray}
 \p^m F_{mn}=0,
\label{maxeuc}
\end{eqnarray}
where $m,n = 1,2,3,4$. We extend the coordinate $x_4$ to a complex
variable $z_4 \equiv x_4 + i x_0$.  As in the Klein-Gordon system,
any solution to the above equations (with sources) can be
extended to a solution of the complexified Maxwell equations by
simply replacing $x_4$ in the solution by $z_4$.  Restricting our
attention to the $x_4=0$ subspace, we obtain the Lorentzian
Maxwell equations
\begin{eqnarray}
 \p^\mu F_{\mu\nu}=0
\label{maxmin}
\end{eqnarray}
with $\mu,\nu=0,1,2,3$. Here note that we replaced the field $A_4$
by $A_0$ by the rule
\begin{eqnarray}
 A_4 = -i A_0,
\end{eqnarray}
which is necessary in accordance to the replacement $\p/\p z_4
\rightarrow -i\p/\p x_0$ which was found in the general observation
in the previous KG system. Due to this replacement, the metric in
(\ref{maxmin}) is $(-1,1,1,1)$ and no $i$ appears in the equation.

To apply this procedure to obtain a new Lorentzian solution from a
Euclidean solution, we have a further requirement that the resultant
Lorentzian solution is real. Note that for this purpose we are
allowed to solve the first Euclidean equation (\ref{maxeuc}) with
complex gauge fields.

\subsubsection{Wick rotation of electrons: cylindrical waves}

Let us present an explicit solution based on an electron-like
solution in the Euclidean space. We adopt an ansatz
$A_1 = A_2 = A_4=0$ for simplicity.
Then the equations of motion of the Euclidean
Maxwell system are just
\begin{eqnarray}
( \p_1^2 + \p_2^2 + \p_4^2) A_3 =0, \quad \p_3 A_3 = 0.
\end{eqnarray}
One example of the solutions is the electron-type,
$A_3=C/\sqrt{x_1^2 + x_2^2 + x_4^2}$,
which is an analog of a static electron potential, where a singular
source is located at $x_1=x_2=x_4=0$.
(The ``real'' electron can be obtained if we Wick rotate $A_3$ to
$iA_0$.) Then the above Wick rotation
procedure gives the following Lorentzian solution
\begin{eqnarray}
 A_3=\frac{C}{\sqrt{x_1^2 + x_2^2 - x_0^2}}, \quad
A_1 = A_2 = A_0=0.
\end{eqnarray}
This is a nontrivial time-dependent solution divergent at $x_0 =
\pm \sqrt{x_1^2 + x_2^2}$. This divergence, which is linearly
aligned along $x_3$, is propagating at the speed of light and
lies on a light cone in the $(x_0,x_1,x_2)$ spacetime.

We consider an imaginary-source extension of this solution to
find nonsingular solutions. One might have already noticed that
the situation is almost the same as the massless KG imaginary
source solution in 1+2 dimensions (note that it is not 1+3
dimensions). After the Wick rotation for a pair of charges
located at $x_4=ix_0=\pm s$, we obtain a Lorentzian solution $A_0
= A_1 = A_3=0$ with $A_3$ identified with $\phi$ in
(\ref{12imag+}). Again from the combination in $\eta$ we expect a
picture of colliding ripples as in the case of the massless KG
system.  Here however the solution is independent of $x_3$ and
thus this is a colliding cylinder elongated along $x_3$ whose
radius is shrinking ($x_0<0$) and growing ($x_0>0$) with the
speed of light.  We have an energy concentration only on this
single cylinder whose radius is changing with speed of light, so
this is a cylindrical ``pulse'', rather than an oscillating wave.
Translational symmetry of this solution along $x_3$ is reflected
in the location of the imaginary sources.  They are at $x_0=\pm i
s$, $x_1=x_2=0$ and the sources are homogeneously distributed
along $x_3$.

Using the gauge field configuration obtained, we may compute the
electric and magnetic fields.  The non-vanishing components are
$E_3$, $B_1$ and $B_2$. The configuration looks quite similar to a
supertube in type II superstring theory --- but in the present
case of course there is no D2-brane to support the cylinder. One
would expect that the cylinder should shrink with the speed of
light, which does coincide with what we have calculated.

Another possibility is that, like the pair of the harmonic
functions in (\ref{12imag+}), we could take opposite charges
instead (or in other words, we could take an imaginary part of
the complex solution for $A_3$). As in Sec.~\ref{timearray}, we
may consider a continuous array of these solutions in time. If we
continuously arrange them in time and thus integrate the
solutions over time, we get a nonzero constant. This is different
from the result of Sec.~\ref{timearray}, but this is still
consistent with the fact that they do solve the equations of
motion.

\subsubsection{Wick rotation of monopoles}

For the case of ``electrons,'' we implicitly made a double Wick
rotation: first, in the usual electron solution we replaced the
electrostatic potential by a Euclidean gauge field $A_3$, and
then we made the Wick rotation for imaginary sources by replacing
$A_4$ by $-iA_0$. We perform a similar analytic continuation for
the Dirac monopoles. We start with embedding a Dirac monopole
configuration into 4 dimensional Euclidean space as
\begin{eqnarray}
A_1 = \frac{-gx_2}{r(r+x_4)},
\quad A_2 = \frac{gx_1}{r(r+x_4)}, \quad
A_3=A_4=0,
\end{eqnarray}
where $r\equiv \sqrt{(x_1)^2+(x_2)^2 + (x_4)^2}$. The embedding
is into the 3-subspace spanned by $(x_1,x_2,x_4)$. Note that the
Dirac string is at the negative axis of $x_4$. Then what we do is
first to shift the origin of the monopole from $(0,0,0)$ to
$(0,0,-s)$, and then to perform the Wick rotation $x_4 \to ix_0$.
After these operations, the Dirac string is located at
$x_1=x_2=0$ and $ix_0<-s$. The important point here is that we
shift the origin (the location of the source) in such a way
that the Dirac string doesn't appear in the real section of
$(x_0,x_1,x_2)$ spacetime.  Taking the singular limit where $s=0$
however does need more than one gauge patch as the singularity
now exists in real spacetime.

The gauge field $A_4$ becomes $-iA_0$ but this is
vanishing from the first place. The nontrivial gauge fields
$A_1$ and $A_2$ become complex, so we add (or subtract)
their complex conjugate to make them real.
Since we succeed in hiding the Dirac string completely in imaginary
spacetime, the resultant configuration is everywhere regular.
For example, we obtain
\begin{eqnarray}
 A_1=\frac{-gx_2}{r(r+s+ix_0)} + {\rm c.c.}
\end{eqnarray}
where now $r= \sqrt{(x_1)^2+(x_2)^2-(x_0)^2 + s^2 + 2isx_0}$.
One can easily see that this is nonsingular, because neither
$r$ nor $r+s+ix_0$ can vanish for nonzero $s$.

Again, the obtained configuration is a cylindrical electro-magnetic
wave (pulse), because the configuration does not depend on $x_3$.

\subsubsection{Wick rotation of instantons: spherical waves}

Next, let us take an Abelian instanton as a starting point. First we
review this singular solutions in 4 dimensional Euclidean space. The
gauge field of the solution is written as
\begin{eqnarray}
 A_m(x) = b_{mn}x_n h(r), \quad
 b \equiv \left(
\begin{array}{cccc}
0&1&0&0 \\[-1.5mm]
-1&0&0&0 \\[-1.5mm]
0&0&0&1\\[-1.5mm]
0&0&-1&0
\end{array}
\right),
\end{eqnarray}
where $r\equiv \sqrt{x_1^2 + x_2^2 + x_3^2 + x_4^2}$.
A straightforward computations show that
\begin{eqnarray}
&& F_{12} = -2h(r) - (x_1^2 + x_2^2) h'(r)/r,
\quad  F_{34} = -2h(r) - (x_3^2 + x_4^2) h'(r)/r,
\nonumber\\
&& F_{13}=F_{24}= (x_1x_4-x_2x_3)h'(r)/r,
\quad F_{23}=F_{41}= (x_2x_4-x_1x_3)h'(r)/r.
\nonumber
\end{eqnarray}
Thus the anti-self-duality equation
$ F_{mn} = -\frac12 \epsilon_{mnpq} F_{pq}$
can be satisfied if we require an equation $F_{12}=-F_{34}$ which is
equivalent to
$-4h(r) = \frac{h'(r)}{r}$.
This can be easily solved as
\begin{eqnarray}
 h(r) = \frac{C}{r^4}.
\end{eqnarray}
This is the Abelian instanton solution with a singularity at $r=0$.

The Wick rotation $x_4\rightarrow i x_0$ procedure gives a
solution
\begin{eqnarray}
 A_1 = x_2 h(r), \quad
 A_2 = -x_1 h(r), \quad
 A_3 = ix_0 h(r), \quad
 A_0 = -ix_3 h(r), \quad
\end{eqnarray}
with the same $h(r)=C/r^4$ but now with $r\equiv \sqrt{x_1^2 +
x_2^2 + x_3^2 - x_0^2}$. In this solution, $A_3$ and $A_0$ are
imaginary. To make a real solution, we may add a complex
conjugate of the solution itself.  When $C$ is real, the result
is,
\begin{eqnarray}
 A_1 = x_2 \frac{2C}{r^4}, \quad
 A_2 = -x_1 \frac{2C}{r^4}, \quad
 A_3 = A_0 = 0. \label{real-inst}
\end{eqnarray}
On the other hand, if we take imaginary $C=i \tilde{C}$ where
$\tilde{C}$ is real, then
\begin{eqnarray}
 A_1 = A_2=0, \quad
 A_3 = -x_0 \frac{2\tilde{C}}{r^4}, \quad
 A_0 = x_3 \frac{2\tilde{C}}{r^4}. \label{imag-inst}
\end{eqnarray}
These two solutions are both time dependent but still singular at
the spacetime origin $r=0$.  While the self-duality relation is
broken when we added the complex conjugate, due to linearity the
Maxwell equations of motion are still satisfied.

Let us move the singularity of the original Euclidean solution a
little bit, as before, to regularize the solution. We start with the
multiple instanton solution
\begin{eqnarray}
 A_m = \sum_{a=1}^N
b_{mn}(x_n - \delta_{n4}s_a) \frac{C_a} {(x_1^2 + x_2^2 + x_3^2 +
(x_4-s_a)^2)^2}.
\end{eqnarray}
The Wick rotation procedure produces a new Lorentzian solution
\begin{eqnarray}
&& A_1 = \sum_{a=1}^N \frac{C_ax_2}
{(x_1^2 + x_2^2 + x_3^2 + (ix_0\!-\!s_a)^2)^2},
\quad
 A_2 = \sum_{a=1}^N \frac{-C_ax_1}
{(x_1^2 + x_2^2 + x_3^2 + (ix_0\!-\!s_a)^2)^2},
\nonumber\\
&& A_3 = \sum_{a=1}^N \frac{C_a(ix_0-s_a)}
{(x_1^2 + x_2^2 + x_3^2 + (ix_0\!-\!s_a)^2)^2},
\quad
 A_0 = \sum_{a=1}^N \frac{-iC_ax_3} {(x_1^2 + x_2^2 + x_3^2 +
(ix_0\!-\!s_a)^2)^2}.
\nonumber
\end{eqnarray}
For simplicity we take a single location of a source ($N=1$), and
add the complex conjugate to make the solution real. Supposing that
$C$ is real, then we obtain a solution
\begin{eqnarray}
&& A_1 = \frac{2 C x_2 (\Delta^2 - 4 s^2 x_0^2)} {(\Delta^2 + 4 s^2
x_0^2)^2}, \quad
 A_2 = \frac{-2 C x_1 (\Delta^2 - 4 s^2 x_0^2)}
{(\Delta^2 + 4 s^2 x_0^2)^2},
\nonumber\\
&& A_3 = \frac{-2 C s (\Delta^2-4s^2x_0^2+4x_0^2 \Delta^2) }
{(\Delta^2 + 4 s^2 x_0^2)^2},
\quad
 A_0 = \frac{8 C sx_0 x_3 \Delta  }
{(\Delta^2 + 4 s^2 x_0^2)^2}, \label{maxsolsph}
\end{eqnarray}
where $\Delta \equiv x_1^2 + x_2^2 + x_3^2 -x_0^2 + s^2$.
It is easy to see that this is a regular configuration, because the
denominator is always positive.

Let us look at a physical property of this time-dependent solution.
The energy concentration of this time-dependent solution can be seen
clearly in its asymptotic expansion in which the parameter $s$ in
$\Delta$ is taken to be negligible, that is, $(-x_0)^2 + (x_1)^2 +
(x_2)^2 + (x_3)^2)^3 \gg s^2$. Physical field strengths have the
following asymptotic values,
\begin{eqnarray}
&& F_{12} = \frac{4C((x_1)^2 + (x_2)^2 -(x_3)^2 + (x_0)^2)}
{((-x_0)^2 \!+\! (x_1)^2 \!+\! (x_2)^2 \!+\! (x_3)^2)^3}, \quad
 F_{13} =
\frac{8C x_2 x_3}
{(-(x_0)^2\! +\! (x_1)^2 \!+\! (x_2)^2 \!+\! (x_3)^2)^3},
\nonumber \\
&& F_{23} = \frac{-8C x_1 x_3}{(-(x_0)^2\! +\! (x_1)^2 \!+\! (x_2)^2
\!+\!
(x_3)^2)^3}, \quad F_{01} = \frac{8C x_2 x_0}{(-(x_0)^2 \!+\!
(x_1)^2 \!+\!
(x_2)^2 \!+\! (x_3)^2)^3}, \nonumber
\\
&& F_{02} = \frac{-8C x_1 x_0}{(-(x_0)^2 \!+\! (x_1)^2 \!+\!
(x_2)^2 \!+\!
(x_3)^2)^3}, \quad F_{03} = 0.
\end{eqnarray}
Note that all the entries have a Lorentz invariant length in their
denominators. Therefore the solution (\ref{maxsolsph}) describes a
2-sphere contracting and expanding with the speed of light --- the
solution is specified by $\Delta$ and if one looks at an equal
$\Delta$ surface one can recognize this geometrical picture.

Let us see what happens at $x_0=0$. In the solution, $A_0$
vanishes, and resultantly the electric field vanishes; there is a
magnetic field, locally concentrated near the origin
$x_1=x_2=x_3=0$. We have plotted some of the gauge field
strengths a little after the collision incident $x_0=0$, in
Fig.~\ref{figf2}. After a long while, the spherical wave (pulse)
expands, as seen in Fig.~\ref{figf4}.

\FIGURE[t]{
\epsfig{file=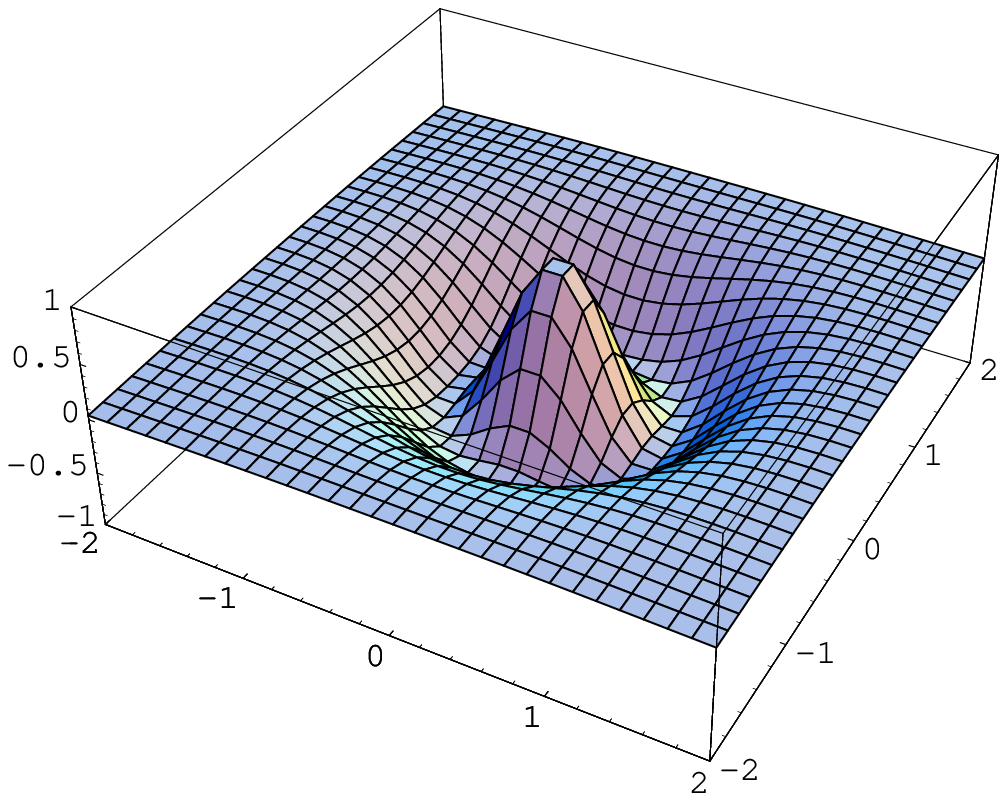,width=6.5cm}
\epsfig{file=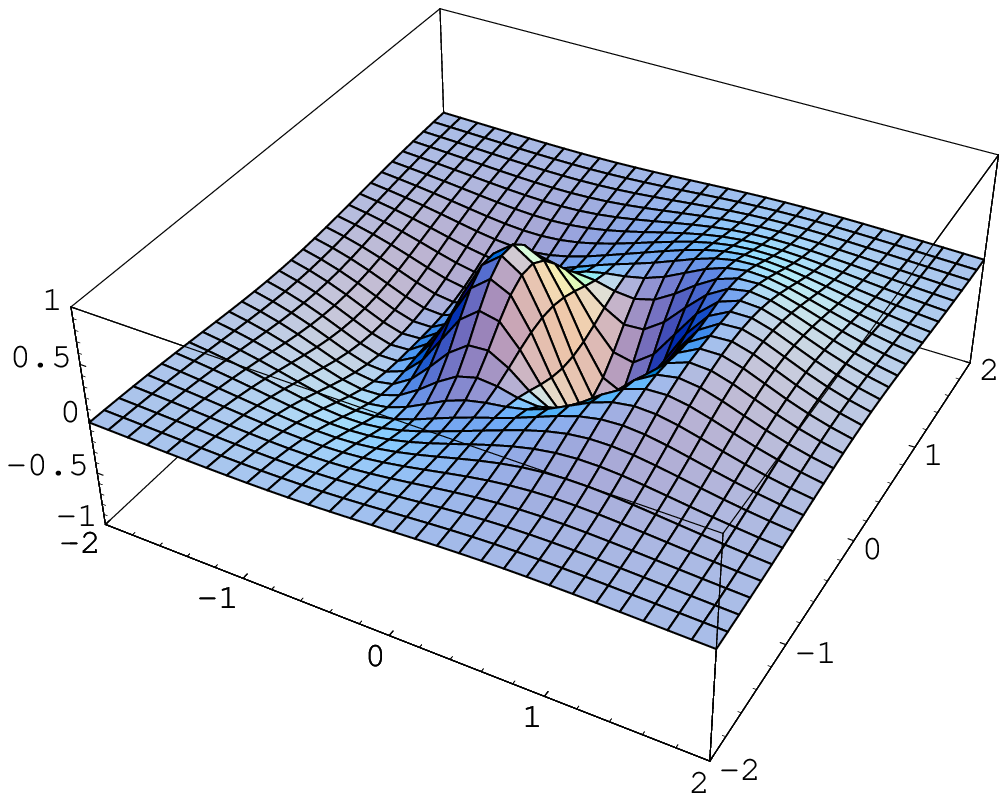,width=6.5cm}
\caption{
Left: the magnetic field $F_{12}$ of the solution
(\ref{maxsolsph}) at a time slice a little after
 the collision $x_0=0$, in a 2 dimensional slice $x_3=0$.
Right: the magnetic field $F_{13}$ at the same time and slice.
This
 component vanishes at $x_0=0$.}
\label{figf2}
}

\FIGURE{
\epsfig{file=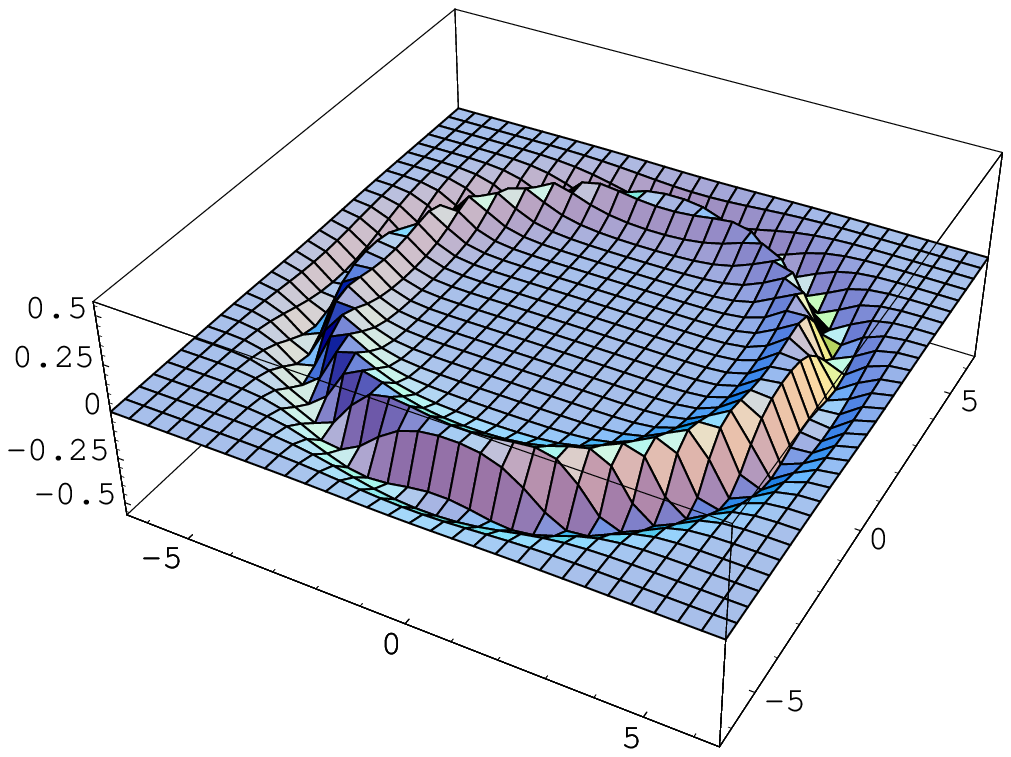,width=6.5cm}
\epsfig{file=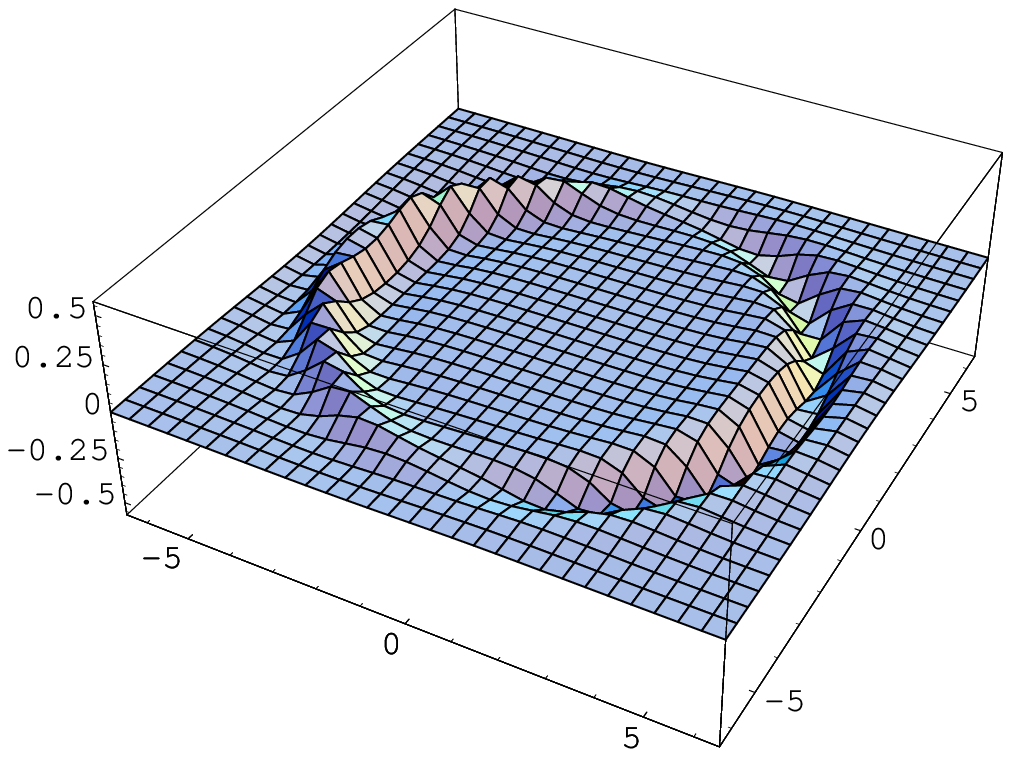,width=6.5cm}
\caption{
Left:
The field strength $F_{12}$ long time after the collision. One
 can see that the spherical wave (pulse) expands.
Right:
The same can be observed for the component $F_{13}$. But
the amplitudes of the waves are small, which is consistent with the
 conservation of the energy.
}
\label{figf4}
}

One may think of $C$ as a pure imaginary constant, before adding
the complex conjugate. Then the resultant solution is different
from (\ref{maxsolsph}). This can be thought of as ``dual'' of the
solution (\ref{maxsolsph}), because at the collision incidence
$x_0=0$, the electric field is nonzero while the magnetic field
vanishes.  In these solutions there is a close and fascinating
relationship between electric magnetic duality and analytic
continuation.  This is seen for example in Eqs.~(\ref{real-inst})
and (\ref{imag-inst}) which can be thought of as related by
analytic continuation.  From a short calculation one finds that
these give rise to field strengths dual to each other. For
example, $F_{03}$ of (\ref{imag-inst}) is equal to $F_{12}$ of
(\ref{real-inst}) if we identify $C=-\tilde{C}$. This is similar
then to the usual case of electric magnetic duality however here
we see that it applies to a new class of solutions coming from
analytic continuations of the instanton.

Of course, $C$ can be an arbitrary complex constant, and thus there
can be a mixture of these electric and magnetic collisions. Infinite
arrays of such Abelian instantons can also be explicitly summed just
like in the Klein-Gordon case using the same hyperbolic tangent
formulas.  The resulting solution is similar to a time dependent
dyon.

\section{Further Generalizations}
In this paper we have developed the utility of complexified
spacetime and its imaginary sources.  While it is unusual to think
of fluctuations of fields to be related to anything else but
physically accessible real spacetime objects, we conclude that
complexification of spacetime is a useful viewpoint. Our main
results include the proof that imaginary source configurations can
be thought to be a new way to understand solutions to homogeneous
differential equations since we are able to prove in several
classes of theories that every smooth solution has at least one
corresponding imaginary source configuration. In mathematical
terms our result can be viewed in the linear cases as unifying the
analysis of singular and non-singular solutions to differential
equations by extending Green's function techniques. Therefore we
expect the methods of imaginary sources to be a useful and
practical tool in many other settings both linear and nonlinear.
The following applications might be interesting,
\begin{itemize}
 \item Non-Abelian gauge theory
\hspace{20mm}$\bullet$ Dirac-Born-Infeld theory
\item Gravity
\hspace{52.5mm} $\bullet$ String theory
\end{itemize}
as all are nonlinear in equations and so the reality condition
will be difficult to be satisfied. But it is possible that in
special examples the reality can be ensured.  For example in the
case of gravity, this should be explicitly possible in all Weyl
type ansatzes.  In string theory, imaginary D-branes in pairs
\cite{sthermo, imagdbrane, greg1} or odd numbers \cite{cards1} do
give rise to sensible backgrounds.  From our discussion it is
clear that imaginary D-branes do not exhaust the physical closed
string backgrounds even at the linear massless level.  To produce
independent dilaton and gauge field configurations, at minimum it
is necessary to use non-BPS imaginary D-branes.  The notion of
imaginary strings must also be developed to produce the source
free background B-fields.  A notion of imaginary strings is in
fact probably already necessary to understand the dynamics of
imaginary D-branes.

To what extent are our current results related to such nonlinear
theories. In this paper we provided proofs of the completeness of
the imaginary sources even for the Maxwell system.
One should note that the Klein-Gordon equation arises in many
important cases such as the linearizations of gravity and string
theory.  To the extent that we trust that the linear
approximations have corresponding nonlinear exact solutions, we
have proved that imaginary sources are relevant in gravity and
string theory.  However this is not so straightforward.  One might
argue that in nonlinear theories such as field theories with
nontrivial interaction terms there will be no notion of ``basis''
because field configurations can not simply be added together to
produce a new configuration.  We hope though that S-charges
associated to given time-dependent solutions (when the
completeness conjecture is correct it always exists) would be an
interesting and intrinsic ways to characterize field
configurations.

In addition to understanding the realm of applicability of the
imaginary source conjecture in other theories, there are many
interesting directions to pursue. For example in gravitational
theories one should develop the notion of ADM S-mass. One can also
try to describe the interaction of imaginary sources with each
other and real sources.  Describing such processes would in fact
lead to an alternate interpretation of positron electron
annihilation.  Such unusual interactions between imaginary
sources and real sources should arise since we have shown
asymptotically they can have the same mass and charge so these
quantum numbers can be conserved in interactions.  A further
natural question is can one generalize the notion of spontaneous
symmetry breaking to the case of imaginary source solutions. This
would include the development of solutions where the delta
functions can be field dependent and take imaginary values in
contrast to delta functions which are just spacetime dependent.
Moreover, the arrays infinitely extended in time give rise to
constant field configurations.  If one could produce similar
imaginary D-branes extended in time, they might be a useful clue
to describing field theory duals to gravity theories in flat
space.  Finally it should be interesting to understand if notions
such as T-duality are applicable to the case of imaginary sources.

\vspace{3mm} \noindent {\bf Acknowledgements} \hspace{10mm}
K.~H.~would like to thank G.~W.~Gibbons for fruitful discussions and
valuable comments. The work of K.~H.~is
supported in part by the National Science Foundation under Grant
No.\ PHY99-07949, by the Japan Society for Promotion of Science, and by
the Royal Society International Grants.
J.~E.~W. is supported in part by the National
Science Council under the NSC grant number NSC 94-2119-M-002-001 and
by the National Center for Theoretical Sciences.

%%%%%%%%%% References %%%%%%%%%%%%%%%%%%%%%%%%%
\newcommand{\J}[4]{{\sl #1} {\bf #2} (#3) #4}
\newcommand{\andJ}[3]{{\bf #1} (#2) #3}
\newcommand{\AP}{Ann.\ Phys.\ (N.Y.)}
\newcommand{\MPL}{Mod.\ Phys.\ Lett.}
\newcommand{\NP}{Nucl.\ Phys.}
\newcommand{\PL}{Phys.\ Lett.}
\newcommand{\PR}{ Phys.\ Rev.}
\newcommand{\PRL}{Phys.\ Rev.\ Lett.}
\newcommand{\PTP}{Prog.\ Theor.\ Phys.}
\newcommand{\hep}[1]{{\tt hep-th/{#1}}}
%%%%%%%%%%%%%%%%%%%%%%%%%%%%%%%%%%%%%%%%%%%%%%%

\end{document}